\documentclass{iopart}

\pdfoutput=1

\usepackage{graphicx,epsfig,sidecap}
\usepackage{bm}

\catcode`@=11 
\renewcommand\tableofcontents{%
  \section*{\contentsname}%
  \@starttoc{toc}%
}
\catcode`@=12

\usepackage{amssymb}

\def\be{\begin{equation}}
\def\ee{\end{equation}}

\def\bea{\begin{eqnarray}}
\def\eea{\end{eqnarray}}

\def\Tr{{\rm Tr}}

\def\e{\epsilon}

\def\a{\alpha}
\def\l{\lambda}
\def\HH{\mathcal H}

\def\t#1{\overline{#1}}
\gdef\Db{{\overline\Delta}}
\newcommand{\tw}{{\cal T}}

\begin{document}

\title[Entanglement negativity in extended systems ]
{Entanglement negativity in extended systems: A field theoretical approach}

\author{Pasquale Calabrese$^1$, John Cardy$^{2}$, and Erik Tonni$^3$}
\address{$^1$ Dipartimento di Fisica dell'Universit\`a di Pisa and INFN,
             Pisa, Italy.\\
         $^2$ Oxford University, Rudolf Peierls Centre for
          Theoretical Physics, 1 Keble Road, Oxford, OX1 3NP, United Kingdom
          and All Souls College, Oxford.\\
          $^3$ SISSA and INFN, via Bonomea 265, 34136 Trieste, Italy. }

\date{\today}

\begin{abstract}

We report on a systematic approach for the calculation of the negativity in the ground state 
of a one-dimensional quantum field theory. 
The partial transpose $\rho_A^{T_2}$ of the reduced density matrix of a subsystem $A=A_1\cup A_2$
is explicitly constructed as an imaginary-time path integral and from this  the 
replicated traces $\Tr (\rho_A^{T_2})^n$ are obtained.
The logarithmic negativity ${\cal E}=\log||\rho_A^{T_2}||$ is then the continuation to $n\to1$ of the 
 traces of the {\it even} powers. 
For pure states, this procedure reproduces the known results. 
We then apply this method to conformally invariant field theories in several different physical situations for 
infinite and finite systems and without or with boundaries. 
In particular, in the case of two adjacent intervals of lengths $\ell_1,\ell_2$ in an infinite system, we derive
the result ${\cal E}\sim(c/4)\ln(\ell_1\ell_2/(\ell_1+\ell_2))$, where $c$ is the central charge. 
For the more complicated case of two disjoint intervals, we show that the negativity  depends only on the harmonic ratio of the four 
end-points and so is manifestly scale invariant. 
We explicitly calculate the scale-invariant functions for the replicated traces in the case
of the CFT for the free compactified boson, but we have not so far been able to obtain the $n\to 1$ continuation for the 
negativity even in the limit of large compactification radius.
We have checked all our findings against exact numerical results for the harmonic chain 
which is described by a non-compactified free boson.

\end{abstract}

\maketitle

\tableofcontents

\section{Introduction}

The study of the entanglement content of many-body quantum systems has allowed in recent 
years a deeper understanding of these systems, in particular in connection with 
criticality and topological order (see \cite{rev} for reviews).
In the case when a system is in its ground-state (or any given pure state) 
the entanglement between two complementary parts is measured by the 
entanglement entropies defined as follows.
Let $\rho$ be the density matrix of a system, which we take to be
in a pure quantum state $|\Psi\rangle$, so that $\rho=|\Psi\rangle\langle\Psi|$. 
Let the Hilbert space be written as a direct product $\HH=\HH_A\otimes\HH_B$. $A$'s reduced density
matrix is $\rho_A=\Tr_B \rho$. The entanglement entropy is the
corresponding von Neumann entropy 
\be 
S_A=-\Tr\, \rho_A \ln \rho_A\,, 
\label{Sdef} 
\ee 
and analogously for $S_B$. When $\rho$
corresponds to a pure quantum state $S_A=S_B$. 
Other standard measures of bipartite entanglement in pure states are  
the R\'enyi entropies 
\be 
S^{(n)}_A= \frac{1}{1-n} \ln {\rm Tr}\,\rho_A^n\,, 
\label{Sndef}
\ee 
that also satisfy $S_A^{(n)}=S_B^{(n)}$ whenever $\rho$ corresponds to a pure quantum state. 
From these definitions $\displaystyle S_A=\lim_{n\to1} S_A^{(n)}$. 
All $S^{(n)}_A$ for any $n$ and for pure states are {\it entanglement monotones} \cite{v-98}, i.e. are quantities which 
do not increase under LOCC (local operation and classical communication), which is a key 
property of any quantity to be a good measure of entanglement.
The knowledge of the R\'enyi entropies for any $n$ determines also 
the full spectrum of the reduced density matrix \cite{cl-08}.

For a mixed state the entanglement entropies are not
 longer  good measures of entanglement since they mix
quantum and classical correlations (e.g. in an high temperature
mixed state, $S_A$ gives the extensive result for the
thermal entropy that has nothing to do with entanglement). 
This is also evident from the fact that $S_A$ is no longer equal to $S_B$.
A quantity that is easily constructed from $S_A^{(n)}$
and $S_B^{(n)}$ is the (R\'enyi) mutual information, defined as 
\be
I^{(n)}_{A:B}=S^{(n)}_A+S^{(n)}_B-S^{(n)}_{A\cup B}\,, 
\ee 
that by definition is symmetric in $A$ and $B$. 
However, $I^{(n)}_{A:B}$ has {\it not} all the correct properties to be an entanglement measure 
and indeed it is not an entanglement monotone for any $n$ (for example it 
has been shown that for most of separable mixed states it is non-zero \cite{pv-07}).

This has also importance for a system in a pure state,  but if one  is 
interested into the entanglement between 
two non-complementary parts $A_1$ and $A_2$.
Indeed, generically the  union $A_1\cup A_2$ is in a mixed state 
with density matrix $\rho_{A_1\cup A_2}=\Tr_B \rho$ with $B$ the complement of 
$A_1\cup A_2$.
As a matter of fact, also the converse is true: any mixed state can be obtained by tracing out some degrees
of freedom from a properly defined larger system, a procedure called purification (see e.g.  \cite{purif}).

A proper definition of the bipartite entanglement for a general mixed state 
(or equivalently the tripartite entanglement in a pure state) has been longly a problem
because most of the proposed measures rely on algorithms rather than explicit expressions 
and are therefore hard to evaluate analytically  (see e.g. Refs. \cite{rev,pv-07,ep-99,varmix}).
However a {\it computable measurement of entanglement},  called {\it negativity}, 
has been introduced 
in a seminal work by Vidal and Werner \cite{vw-01}. They also showed 
how the negativity provides bounds on a few operationally well-defined measures of
entanglement (such as distillable entanglement or teleportation fidelity). 
The precise meaning of the negativity in quantum information has been indeed established in Ref. 
\cite{vw-01}, but the main reason for its success is a practical one:
the negativity is obtained  from any mixed state by computing a partial
transposition and then diagonalizing a matrix, while 
we still do not know  generically how to compute
any other mixed-state entanglement measures.

Following Ref. \cite{vw-01},  the negativity is defined as follows.
Let us consider a density matrix $\rho$ corresponding to a given 
mixed or pure state acting on a Hilbert space $\HH$. 
Let us consider the bipartition $\HH={\cal H}_1\otimes {\cal H}_2$ and let us
denote by $|e_i^{(1)}\rangle$ and $|e_j^{(2)}\rangle$ two arbitrary bases 
in the Hilbert spaces  of each part. 
The partial transpose (e.g. with respect to the second space) of  $\rho$ is defined as
\be 
\langle e_i^{(1)} e_j^{(2)}|\rho^{T_2}|e_k^{(1)} e_l^{(2)}\rangle=\langle e_i^{(1)} e_l^{(2)}|\rho| e^{(1)}_k e^{(2)}_j\rangle,
\ee
and then the {\it logarithmic} negativity as
\be
{\cal E}\equiv\ln ||\rho^{T_2}||=\ln \Tr |\rho^{T_2}|\,,
\ee
where the trace norm  $||\rho^{T_2}||$ is
the sum of the absolute values of the eigenvalues $\lambda_i$ of $\rho^{T_2}$.
In Ref. \cite{vw-01} another measure, termed simply negativity, has been also introduced
\be
{\cal N}\equiv \frac{||\rho^{T_2}||-1}{2}\,,
\ee
which is trivially related to ${\cal E}$ as ${\cal N}=(e^{\cal E}-1)/2$.
However,  ${\cal E}$ is additive, while  ${\cal N}$ is not and for this reason 
we will concentrate in the following on ${\cal E}$ from which ${\cal N}$ can be trivially derived.
When the two parts are  two microscopic degrees of freedom (e.g. spins),
the  negativity is equivalent to other commonly used entanglement estimators 
such as the concurrence \cite{rev,fazio}.
However the definition of the negativity  is more appealing because it is basis independent and so calculable 
by quantum field theory (QFT) which 
naturally unveils  universal features, in particular close to a quantum critical point. 
For 1D critical theories, that at low energy are also Lorentz invariant, 
the powerful tools of conformal field theory (CFT) can be applied. 

The fact that the negativity is computable made it 
a remarkable tool to  study the tripartite entanglement content of many 
body quantum systems both in their ground-state 
\cite{Audenaert02,Neg1,Neg2,Neg3,sod2,sod1} or out of equilibrium \cite{sod4,sod3}.
Some studies for the bipartite entanglement at finite temperature have been presented as well \cite{finT2,finT1,finT3}. 
However, only in a recent short communication \cite{us-letter}, we carried out a more 
systematic and generic approach to negativity  based on QFT (and in particular  CFT). 
In the following we give detailed derivations of all results announced in Ref.
 \cite{us-letter} and report some other new findings.


\begin{figure}[t]
\includegraphics[width=.9\textwidth]{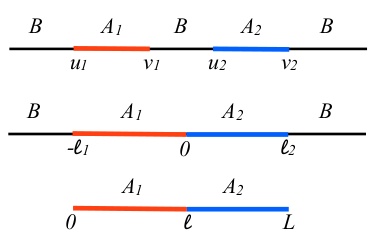}
\caption{The three main configurations of 1D systems we consider.
Top: the entanglement between two disjoint intervals $A_1$ and $A_2$ embedded 
in the ground-state of a larger system formed by the union of $A_1$, $A_2$ and the remainder $B$. 
The whole system can be either finite or infinite.
Middle: The entanglement between two adjacent intervals in a larger system.
Bottom: The entanglement between two adjacent intervals which form the full system ($B\to\emptyset$).
In all cases we denote $A=A_1\cup A_2$.
}
\label{intervals}
\end{figure}

\subsection{Setup and notations.}

In this manuscript we only consider the entanglement, measured by the negativity, in the ground state of 
one-dimensional systems (1D), although some derivations have a more general validity.  
We focus on the tripartitions depicted in Fig. \ref{intervals}, i.e. the entire system is divided 
in three parts $A_1$, $A_2$ and $B$ and we consider the entanglement between $A_1$ and $A_2$,
whose union is generically denoted with $A$.
As usual, $A$'s reduced density matrix is $\rho_A=\Tr_B \rho$.
In order to lighten the notation, we will denote as $\rho_1$ and $\rho_2$ the reduced density 
matrix corresponding to $A_1$ and $A_2$ respectively, i.e.
\bea
\rho_1&\equiv&\rho_{A_1}=\Tr_{A_2}(\rho_A)=\Tr_{B\cup A_2} (\rho)\,,\\ 
\rho_2&\equiv&\rho_{A_2}=\Tr_{A_1}(\rho_A)=\Tr_{B\cup A_1} (\rho)\,.
\eea
Also the partial transpose with respect to $A_1$ and $A_2$ degrees of freedom 
will simply be denoted with the superscripts $T_1$ and $T_2$ respectively, i.e.
\be
\rho_A^{T_2}\equiv \rho_A^{T_{A_2}}\,, \qquad {\rm and}\qquad 
\rho_A^{T_1}\equiv \rho_A^{T_{A_1}}\,.
\ee

\section{A replica approach for the negativity}

\subsection{Replicas and entanglement entropy.}

One of the most successful approaches to calculate the entanglement entropy for a bipartite system
is  
based on the replica trick \cite{cc-04} which proceeds as follows. 
One first calculates the traces of integer powers of the reduced density matrix 
\be
{\rm Tr}\rho_2^n=\sum_i\zeta_i^n,
\ee
where  $\zeta_i$ are the eigenvalues of the reduced density matrix  $\rho_2$.
Then, if one is able to analytically continue this expression to general complex $n$,
the entanglement entropy is given by
\begin{equation}
S_{A_2}=-\lim_{n\to1}{\partial\over\partial n}{\rm Tr}\,\rho_2^n
= \lim_{n\to1}\frac1{1-n} \ln {\rm Tr}\,\rho_2^n\,.
\label{SA3}
\end{equation} 
It has been shown \cite{cc-04,cc-rev} that, for integer $n$,  ${\rm Tr}\rho_2^n$ is a partition function on a
complicated Riemann surface (or equivalently the correlation
function of specific {\it twist fields}, as we shall review later) that is analytically
achievable in a quantum field theory.
The analytical properties of ${\rm Tr}\,\rho_2^n$ in the complex plane of replicas are 
also discussed in Refs. \cite{ccd-08,gt-11}.

\subsection{Replicas and negativity.}

A natural way to use a replica trick to compute the negativity would be to relate it to 
the traces of integer powers of $\rho^{T_2}$. 
The trace norm of $\rho^{T_2}$ can be written in terms of its 
eigenvalues  $\lambda_i$ as
\be
 \Tr |\rho^{T_2}|= \sum_i |\lambda_i|= \sum_{\l_i>0} |\l_i|+ \sum_{\l_i<0} |\l_i|=1+ 2  \sum_{\l_i<0} |\l_i|,
\ee
where in the last equality we used the normalization $\sum_i \lambda_i=1$.
This expression makes evident that the negativity measures ``how much'' the eigenvalues 
of the partial transpose of the density matrix are negative, a properties which is the reason of 
the name negativity. 

The traces $\Tr (\rho^{T_2})^n$ of integer powers of $\rho^{T_2}$ have a difference dependence on $|\lambda_i|$
depending on the parity of $n$.
In the following we will always indicate an even $n=2m$ as $n_e$ while an odd one  
$n=2m+1$ with $n_o$. It is understood that $n_e$ and $n_o$ refer always to the 
same thing, it is just the functional dependence of the traces which is different.  
In fact, for $n$ even and odd, the traces of integer powers of $\rho^{T_2}$ are
\bea
\Tr (\rho^{T_2})^{n_e}&=&\sum_i \lambda_i^{n_e}= \sum_{\l_i>0} |\l_i|^{n_e}+ \sum_{\l_i<0} |\l_i|^{n_e}\,, 
\label{trne}\\
\Tr (\rho^{T_2})^{n_o}&=&\sum_i \lambda_i^{n_o}= \sum_{\l_i>0} |\l_i|^{n_o}- \sum_{\l_i<0} |\l_i|^{n_o}\,.
\label{trno}
\eea
If now we just set $n_e=1$ in Eq. (\ref{trne}) we formally obtain $ \Tr |\rho^{T_2}|$ in which we are 
interested. Oppositely, setting $n_o=1$ in Eq. (\ref{trno}) gives the normalization $\Tr \rho^{T_2}=1$.
This means that the analytic continuations from even and odd $n$ are different and 
 the trace norm in which we are interested is obtained
by considering the analytic continuation of the even sequence at $n_e\to1$, 
i.e. 
\be  {\cal E}=\lim_{n_e\to1} \ln \Tr (\rho^{T_2})^{n_e}\,.
\ee
We should also mention that being $(\rho^{T_2})^T=\rho^{T_1}$, 
it trivially holds $\Tr (\rho^{T_2})^{n}=\Tr (\rho^{T_1})^{n}$ for any $n$, even not integer.

At first, this approach can seem rather unnatural because what we are doing is basically 
propose to calculate a quantity for even numbers and at the end set this even number to $1$
which is instead odd.
However, similar replica calculations have been already successfully applied to 
other different physical problems \cite{k-91,otherrep}, showing the reasonableness  
of the approach.

\subsection{Integer powers of the partial transpose and the negativity of a pure state.}
\label{secpure}

As a first example and check of the replica trick, we consider  the case of a bipartition 
of the Hilbert spaces  ${\cal H}={\cal H}_1\otimes {\cal H}_2$ of a pure state 
$|\Psi\rangle$ with  $\rho=|\Psi\rangle\langle\Psi|$ for which the negativity is known \cite{vw-01}.
The Schmidt decomposition, in terms of two bases $|e_k^{(1)}\rangle$ and $|e_j^{(2)}\rangle$  in the 
Hilbert spaces  ${\cal H}_1$ and ${\cal H}_2$ respectively, gives
\begin{equation}
|\Psi \rangle = \sum_{j} c_j \, | e^{(1)}_j   e^{(2)}_j  \rangle 
\end{equation}  
where the coefficient $c_j$ can be chosen such that $c_j\in [0,1]$. We also have
\be\fl
\rho= \sum_{j,k} c_j c_k | e^{(1)}_j  e^{(2)}_j  \rangle  \langle e^{(1)}_k  e^{(2)}_k  |\,,
\qquad 
\rho_2=\Tr_1(\rho)= \sum_j c_k^2 |e^{(2)}_k  \rangle  \langle e^{(2)}_k|\,,
\ee
where $\rho_2$ is the reduced density matrix on ${\cal H}_2$. 

The partial transpose of the density matrix is then 
\begin{equation}
\rho^{T_2}
= 
\sum_{j,k} c_j c_k
| e^{(1)}_j  e^{(2)}_k  \rangle  \langle e^{(1)}_k  e^{(2)}_j  |\,,
\end{equation}  
whose $n$-th power reads
\begin{eqnarray}
\fl (\rho^{T_2})^{n}
& =& 
 \sum_{\tiny \begin{array}{c}
j_1 , \dots , j_n \\ 
k_1 , \dots , k_n
\end{array}}
 c_{j_1} c_{k_1}  c_{j_2} c_{k_2} \dots  c_{j_n} c_{k_n}
\nonumber \\ \fl & & \qquad \times\,
| e^{(2)}_{k_1}  e^{(1)}_{j_1} \rangle  \langle e^{(2)}_{j_1}  e^{(1)}_{k_1}  |
 e^{(2)}_{k_2}  e^{(1)}_{j_2} \rangle  \langle e^{(2)}_{j_2}  e^{(1)}_{k_2}  |
\dots
| e^{(2)}_{k_n}  e^{(1)}_{j_n} \rangle  \langle e^{(2)}_{j_n}  e^{(1)}_{k_n}  |=
\nonumber \\ \fl 
&   =& 
 \sum_{\tiny \begin{array}{c}
j_1 , \dots , j_n \\ 
k_1 , \dots , k_n
\end{array}}
c_{j_1} c_{k_1}  c_{j_2} c_{k_2} \dots  c_{j_n} c_{k_n} 
\nonumber \\ \fl & & \qquad \times\,
| e^{(2)}_{k_1}  e^{(1)}_{j_1} \rangle \,\delta_{j_1,k_2} \delta_{k_1,j_2} 
\,\delta_{j_2,k_3} \delta_{k_2,j_3} 
\dots
\delta_{j_{n-1},k_n} \delta_{k_{n-1},j_n} 
\langle e^{(2)}_{j_n}  e^{(1)}_{k_n}  |\,.
\end{eqnarray}
The sequence of deltas gives a result which depends on the parity of $n$ 
\begin{equation}
\label{trace rhoTA nth pure}
(\rho^{T_2})^{n} =
\left\{
\begin{array}{ll}\displaystyle
\sum_{j_1,k_1} c_{j_1}^{n_o}  c_{k_1}^{n_o} 
 | e^{(2)}_{k_1}  e^{(1)}_{j_1} \rangle \langle e^{(2)}_{j_1}  e^{(1)}_{k_1}  |,
&\qquad n= n_o\; {\rm odd}, 
\\
\displaystyle
\sum_{j_1,k_1} c_{j_1}^{n_e}  c_{k_1}^{n_e} 
\, | e^{(2)}_{k_1}  e^{(1)}_{j_1} \rangle \langle e^{(2)}_{k_1}  e^{(1)}_{j_1}  |,
&
\qquad  n=n_e\; {\rm even},
\end{array}
\right.
\end{equation}  
from which
\be
\Tr (\rho^{T_2})^{n}=
\left\{
\begin{array}{ll}\displaystyle 
 \sum_r c_r^{2 n_o}=
\Tr \rho_2^{n_o} ,
&\hspace{.6cm} n= n_o\; {\rm odd},
\\ \displaystyle 
\bigg[\sum_r c_r^{n_e} \bigg]^2
=
\big( \Tr \,\rho_2^{n_e/2}  \big)^2, 
&
\hspace{.6cm}  n=n_e\; {\rm even}.
\end{array}\right.
\label{pureqm}
\ee
Notice in particular that 
\be
\Tr(\rho^{T_2})^2=\Tr(\rho^{T_2})=1\,.
\ee

Taking the limit  $n_e\to 1$, we recover the result \cite{vw-01}
that for a pure state the logarithmic negativity is the R\'enyi entropy of order $1/2$
(cf. Eq. (\ref{Sndef}) with $n=1/2$)
\be
{\cal E}= S^{(1/2)}_{A_2}=2\ln\Tr\rho_2^{1/2}. 
\label{genEpure}
\ee
Taking instead the limit $n_o\to 1$, we recover the normalization 
$\textrm{Tr} (\rho^{T_2})= \textrm{Tr} \rho_2=1$.

\section{Negativity and Quantum Field Theory}

In this section we show how to compute for a generic tripartition of
a 1D quantum field theory the integer traces of the partial transpose of the 
reduced density matrix and from these, via the replica trick, the logarithmic negativity.  
For concreteness  we will only consider the {\it tripartition} of a 1D system depicted in Fig.~\ref{intervals} with 
$A$ composed of two parts $A=A_1\cup A_2=[u_1,v_1]\cup [u_2,v_2]$
and $B$ the remainder, but most of the following ideas apply to more general cases 
(e.g. $A_1$ and $A_2$ made each of several disjoint intervals).  
In order to introduce the general formalism for the negativity, we first review in the next subsection 
the path integral approach to the entanglement entropy  \cite{cc-04,cc-rev} and the use of the 
twist fields \cite{cc-04,ccd-08}.

\subsection{The reduced density matrix and the entanglement entropy in QFT.}

\begin{figure}[t]
\includegraphics[width=\textwidth]{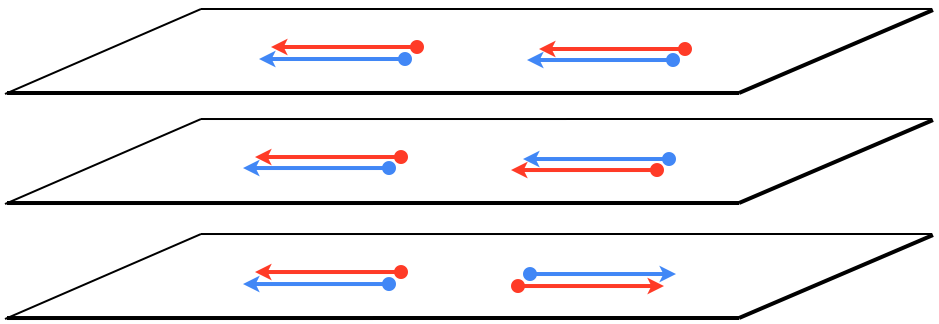}
\caption{Top: The reduced density matrix $\rho_A$ o f two disjoint intervals. 
Middle: Partial transpose with respect to the second interval $\rho_A^{T_{2}}$.
Bottom:  Reversed partial transpose $\rho_A^{C_{2}}=C \rho_A^{T_{2}}C$,
where $C$ reverses the order of the row and column indices.
}
\label{rhos}
\end{figure}

The density matrix $\rho$ in a thermal state at temperature $T=1/\beta$ may be written as a 
path integral in the imaginary time interval $(0,\beta)$
\begin{equation}
\label{pathi}
\fl \rho(\{\phi_x\}|\{\phi'_{x'}\})=
Z^{-1}\int[d\phi(y,\tau)]
\prod_x\delta(\phi(y,0)-\phi'_{ x'})\prod_x
\delta(\phi(y,\beta)-\phi_x)\,e^{-S_E}\,,
\end{equation}
where $Z={\rm Tr}\,e^{-H/T}$ is the partition function and  $S_E$ is 
the euclidean action. 
Here the rows and columns of the density matrix are labelled by the 
values of the fields $\{\phi_x\}$ at $\tau=0,\beta$.
The normalization factor $Z$ ensures ${\rm Tr}\rho=1$, and is found by 
setting $\{\phi_x\}=\{\phi'_{x}\}$ and integrating over these variables. 
In the path integral, this has the effect of sewing together the two edges 
 to form a cylinder of circumference $\beta$.

Now let us consider the subsystem $A$ in Fig.~\ref{intervals}
composed of two parts $A=A_1\cup A_2=[u_1,v_1]\cup [u_2,v_2]$.
The reduced density matrix $\rho_A$ is obtained from (\ref{pathi})
by sewing together only those points $x$ which are not in $A$. This
has the effect of leaving two open cuts, one for each interval
$(u_j,v_j)$, along the line $\tau=0$.
In the limit of zero temperature, i.e. for the ground-state of the QFT, 
the cylinder becomes a plane as  in Fig.~\ref{rhos} (top) where the 
two open cuts correspond to the rows and columns of $\rho_A$
and the orientation of the arrows gives the ordering of the row/column indices,
e.g. increasing along the directions of the arrows.

\begin{figure}[t]
\includegraphics[width=.88\textwidth]{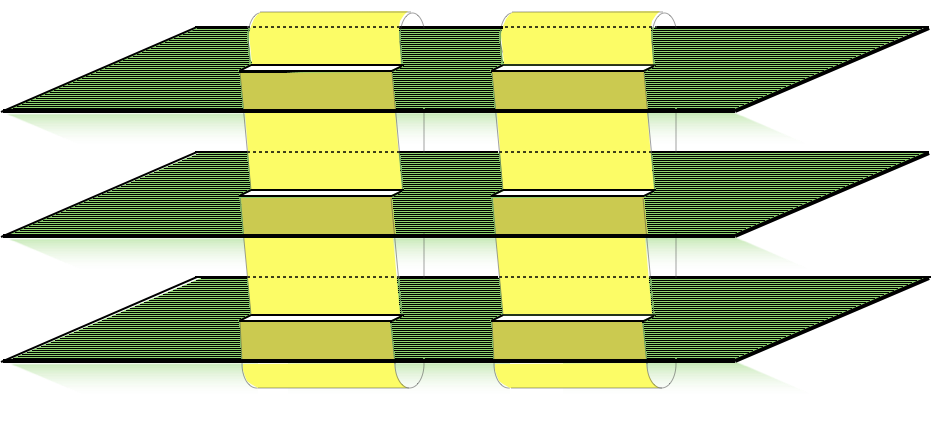}
\caption{The path integral representation of $\Tr\rho_A^n$ gives a $n$-sheeted Riemann surface ${\cal R}_n$
depicted here for $n=3$ and $A=[u_1,v_1]\cup [u_2,v_2]$.
}
\label{replicas}
\end{figure}

We may then compute ${\rm Tr}\,\rho_A^n$, for any positive
integer $n$, by making $n$ copies of the above, labelled by an integer
$j$ with $1\leq j\leq n$, and sewing them together cyclically along the 
the cuts so that $\phi_j(x,\tau=0^-)=\phi_{j+1}(x,\tau=0^+)$ and
$\phi_n(x,\tau=0^-)=\phi_1(x,\tau=0^+)$ for all $x\in A$. 
This defines the $n$-sheeted Riemann surface ${\cal R}_n$ depicted for $n=3$ 
in Fig. \ref{replicas}.
Denoting with $Z_{{\cal R}_n}$ the partition function on this surface we have
\begin{equation}
\label{ZoverZ}
{\rm Tr}\,\rho_A^n=\frac{Z_{{\cal R}_n}}{ Z^n}\,.
\end{equation}
This expression gives the R\'enyi entropies in Eq. (\ref{Sndef}) for integer $n$
and through the analytic continuation the entanglement entropy in Eq. (\ref{Sdef}).

As a consequence of its locality, the
partition function can be expressed as an object calculated
on the complex plane $\mathbb C$, where the structure of the Riemann surface 
is implemented through appropriate boundary conditions around the
points with non-zero curvature, i.e. the partition function in a theory 
defined on the complex plane $z=x+i\tau$
should be written in terms of certain ``fields'' at $z=v_j$ and $z=u_j$.
The partition function (here  ${\cal L}[\phi](z,{\bar z})$ is the 
 lagrangian density)
\be
\label{partfunct}
Z_{{\cal R}_{n}}= \int [d\phi]_{{\cal R}_{n}} 
\exp\left[-\int_{{\cal R}_{n}} d z d{\bar z}\,{\cal L}[\phi](z,{\bar z})\right]\,,
\ee 
essentially defines these fields. 
Following Ref. \cite{ccd-08}, it is useful to move
the topology of the world-sheet (i.e. the space where the coordinates
$x,\tau$ lie) ${\cal R}_{n}$ to the target space (i.e. the space where the 
fields lie).
To this aim, let us consider a model formed by $n$ independent copies of the original 
model.
The partition function  (\ref{partfunct}) can be re-written as
the path integral on the complex plane
\be
\label{partfunctmulti}
\fl Z_{{\cal R}_{n}} =
\int_{{\cal C}_{u_j,v_j}} \,[d\phi_1 \cdots d\phi_n]_{\mathbb C} 
\exp\left[-\int_{\mathbb C} 
d z d {\bar z}\,({\cal L}[\phi_1](z,{\bar z})+\ldots
+{\cal L}[\phi_n](z,{\bar z}))\right],
\ee
where with $\int_{{\cal C}_{u_j,v_j}}$ we indicated the 
{\it restricted}  path integral with conditions
\be
\fl    
{\cal C}_{u_j,v_j}:\quad
\phi_i(x,0^+) = \phi_{i+1}(x,0^-)~,\quad 
x\in [u_1,v_1]\cup  [u_2,v_2] ,\quad i=1,\ldots,n\,,
\label{Cdef}
\ee 
where $n+i\equiv i$. The lagrangian density of the
multi-copy model is
\be
{\cal L}^{(n)}[\phi_1,\ldots,\phi_n](x,\tau) = 
{\cal L}[\phi_1](x,\tau)+\ldots+{\cal L}[\phi_n](x,\tau),
\ee
so that the energy density is the sum of the energy
densities of the $n$ individual copies. Hence the expression
(\ref{partfunctmulti}) indeed defines local fields at
$(u_j,0)$ and $(v_j,0)$ in the multi-copy model \cite{ccd-08}.

The local fields defined in (\ref{partfunctmulti}) are examples of 
{\it twist fields}. 
Twist fields exist in a QFT whenever there is a global internal symmetry 
$\sigma$, i.e.  
$\int dx d\tau {\cal L}[\sigma\phi](x,\tau) = \int dx d\tau {\cal L}[\phi](x,\tau)$. 
The twist fields defined by (\ref{partfunctmulti}),  called {\em branch-point twist fields} \cite{ccd-08}, are 
associated to the two opposite cyclic permutation symmetries $i\mapsto i+1$ and $i+1\mapsto i$. 
We can denote them simply by $\tw_n$ and $\t\tw_n$
\bea 
\tw_n\equiv\tw_\sigma~,\quad &\sigma&\;:\; i\mapsto i+1 \ {\rm mod} \,n\,, \\
\t\tw_n\equiv\tw_{\sigma^{-1}}~,\quad &\sigma^{-1}&\;:\; i+1\mapsto i \ {\rm mod} \,n\,.
\eea
$\t\tw_n$ can be identified with $\tw_{-n}$ and $\tw_n^n=\t\tw_n^n=1$.
Thus, for the $n$-sheeted Riemann surface along the set $A$ made of 
the two disjoint intervals $[u_1,v_1]\cup[u_2,v_2]$, we have
\be
\Tr\rho_A^n=\langle {\cal T}_n(u_1)\overline{\cal T}_n(v_1) {\cal T}_n(u_2)\overline{\cal T}_n(v_2)\rangle_{\mathbb C}\,.
\label{rhoatw}
\ee
In the following the subscript ${\mathbb C}$ will be understood in the expectation values, 
if not differently stated.

\subsection{The partial transposition and the negativity in QFT.}

\begin{figure}[t]
\includegraphics[width=.88\textwidth]{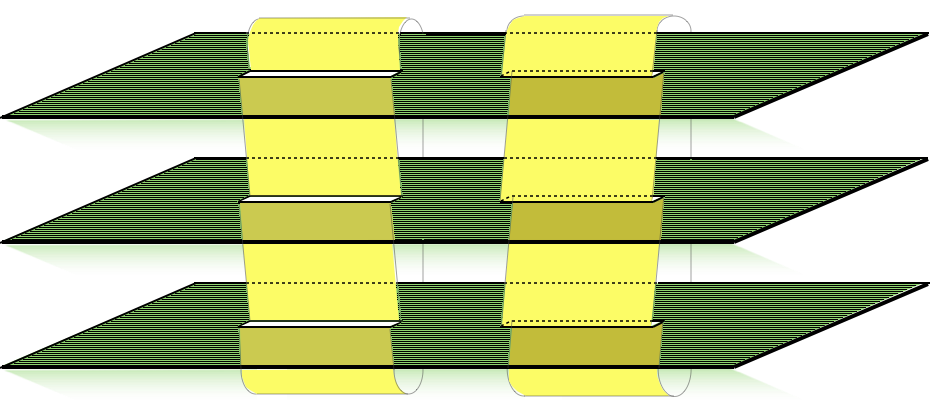}
\caption{Path integral representation of  $\Tr(\rho_A^{T_{2}})^n=\Tr(\rho_A^{C_{2}})^n$ for $n=3$.
}
\label{replicas1}
\end{figure}

The partial transposition of the reduced density matrix $\rho_A$ with respect to  the second interval $A_2$ 
corresponds to the exchange of  row and column indices in $A_2$.  
In the path integral representation, this is equivalent to interchange the upper and lower edges of the second cut
in $\rho_A$ as depicted in the middle of Fig.~\ref{rhos}.
If we join $n$ copies of $\rho_A^{T_2}$ cyclically, we have an $n$-sheeted Riemann
surface where row and column indices are reversed compared to those of a correlation 
function of four twist fields, as it should be clear from the middle of Fig. \ref{rhos}.
This problem can be however solved very easily by
reversing the order of the column and row indices in $A_2$
as in the bottom of Fig.~\ref{rhos}, to obtain the {\it reversed partial transpose} $\rho_A^{C_{2}}$.
This is related to the partial transpose as $\rho_A^{C_{2}}=C \rho_A^{T_{2}} C$, 
where $C$ reverses the order of indices either on the lower or on the upper cut and satisfies $C^2=1$. 
Clearly 
$\Tr(\rho_A^{T_{2}})^n=\Tr(\rho_A^{C_{2}})^n$ 
and so $\Tr(\rho_A^{T_{2}})^n$ is the partition function on the $n$-sheeted surface obtained by joining 
cyclically $n$ of the above $\rho_A^{C_{2}}$ as in Fig.~\ref{replicas1}.
In this case, the order of the row and column indices is the right one to identify this
partition function with the four-point function of the twist fields
\be
\Tr(\rho_A^{T_{2}})^n=\Tr(\rho_A^{C_{2}})^n=
\langle {\cal T}_n(u_1)\overline{\cal T}_n(v_1) \overline{\cal T}_n(u_2){\cal T}_n(v_2)\rangle\,,
\label{4ptdef}
\ee
i.e. the partial transposition has the net effect to exchange two twist operators compared to Eq. (\ref{rhoatw}).
We notice that we could easily have worked out $\Tr(\rho_A^{T_{2}})^n$ without introducing the 
reverse partial transpose. However this is a very useful technical concept because it allows to identify 
$\Tr(\rho_A^{T_{2}})^n$ with the correlation function of already known and studied twist fields, 
without the need of introducing new fields.

For $n=2$, ${\cal T}_2=\overline{\cal T}_2$ and so 
\be 
\Tr\rho_A^2=\Tr(\rho_A^{T_{2}})^2\,,
\label{rhoT2vsrho2}
\ee 
which also straightforwardly follows from the properties of the trace and so it is true for any matrix $\rho$ replacing 
$\rho_A$ above.

To replace $\rho_A^{T_{2}}$ with $\rho_A^{C_{2}}$ it has been fundamental to 
consider integer cyclical traces. The operator $C$ enters in quantities like  
$\Tr(\rho_A\rho_A^{T_{2}})$ 
which is in fact the partition function on a non-orientable surface with the topology of a Klein bottle.
This can also be computed using CFT methods \cite{klein}.

\subsection{The case of two adjacent intervals.}

Eq. (\ref{4ptdef}) is of general validity, but it has interesting simple and general consequences when 
specialized to the case of two adjacent intervals.
This can be obtained by letting $v_1\to u_2$ so that 
\be
\Tr(\rho_A^{T_{2}})^n=
\langle {\cal T}_n(u_1) \overline{\cal T}_n^2(u_2){\cal T}_n(v_2)\rangle\,.
\label{3ptdef}
\ee
For arbitrary $n$, this expression cannot be evaluated in general,
but two cases are easily worked out.

For $n=2$, we have ${\cal T}_2^2=1$ and so 
\be
\Tr(\rho_A^{T_{2}})^2=
\langle {\cal T}_2(u_1){\cal T}_2(v_2)\rangle= \Tr \rho_{A_1\cup A_2}^2\,.
\label{3ptn2}
\ee
This relation has been derived here from QFT methods and it would 
be interesting to know its general validity. 

For $n=3$, we have $ \overline{\cal T}_3^2={\cal T}_3$ and so 
\be
\Tr(\rho_A^{T_{2}})^3=
\langle {\cal T}_3(u_1) {\cal T}_3(u_2){\cal T}_3(v_2)\rangle\,.
\ee

\subsection{The case of a single interval.}

\begin{figure}[t]
\includegraphics[width=.74\textwidth]{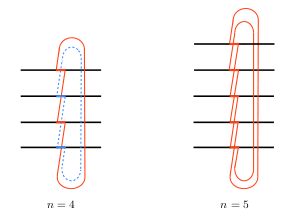}
\caption{Cut of the Riemann surface defined by $\langle {\cal T}^2_{n}(u_2) \overline{\cal T}^2_{n}(v_2)\rangle$.
Left: A typical example for $n$ even ($n=4$ in the figure) showing how the surface decouples in two independent parts, each 
with half of the sheets.
Right: A typical example for $n$ odd ($n=5$ in the figure) showing that the net effect is just a reshuffling of the sheet 
numeration going from $(1,2,3,4,5)$ to $(1,3,5,2,4)$.
}
\label{evenodd}
\end{figure}

We now specialize to a pure state by letting $B\to \emptyset$ for which 
$\Tr (\rho_A^{T_2})^{n}$ can be worked out in full generality 
 considering $u_2\to v_1$ and $v_2\to u_1$ 
\be
\Tr (\rho_A^{T_2})^{n}=\langle {\cal T}^2_{n}(u_2) \overline{\cal T}^2_{n}(v_2)\rangle\,.
\ee
This expression depends on the parity of $n$ because  ${\cal T}_n^2$ connects the $j$-th sheet 
with the $(j+2)$-th one. 
For $n=n_e$ even, 
the $n_e$-sheeted Riemann surface decouples in two independent ($n_e/2$)-sheeted 
surfaces characterized by the parity of the sheets.
Conversely for $n=n_o$ odd, the surface remains a $n_o$-sheeted Riemann
surface (a part from a shuffling of the sheets). 
This is pictorially shown in Fig. \ref{evenodd}.
Thus we have
\bea
\Tr (\rho_A^{T_2})^{n_e}&=& (\langle {\cal T}_{n_e/2}(u_2) \overline{\cal T}_{n_e/2}(v_2)\rangle)^2=
(\Tr\rho_{A_2}^{n_e/2})^2\,,\nonumber \\
\Tr (\rho_A^{T_2})^{n_o}&= & \langle {\cal T}_{n_o}(u_2) \overline{\cal T}_{n_o}(v_2)\rangle=\Tr \rho_{A_2}^{n_o}\,,
\label{pureqft}
\eea
which are the results for pure states in Eq. (\ref{pureqm}), recovered here purely from QFT
and showing in a first simple example the effect of the parity of $n$ on the Riemann surfaces.

\section{Negativity and conformal invariance: general results}

In this section we explicitly calculate the negativity for those situations in CFT where  
$\Tr (\rho^{T_2}_A)^n$ can be fixed by global conformal invariance and  so does not depend on the
operator content of the theory, but only on the central charge. 

Before embarking in the full negativity calculation, we should recall that 
$\Tr \rho_A^n$, for the case of a single interval $A=[u_1,v_1]$, has been obtained by uniformizing the
$n$-sheeted Riemann surface to the complex plane  \cite{cc-04}.  
This shows that the twist fields transform like primary operators of dimension  \cite{cc-04}
\be 
\Delta_{{\cal T}_n}={\Delta}_{\overline{\cal T}_n}=\frac{c}{12} \Big(n-\frac1n\Big),
\ee
where $c$ is the central charge. This implies ($\ell=|u_1-v_1|$) \cite{Holzhey,Vidal,cc-04}
\be
\Tr \rho_A^n=c_n \Big(\frac\ell{a}\Big)^{-{c}/6 (n-1/n)}
\quad\Rightarrow\;\; S_A=\frac{c}3 \ln \frac{\ell}a+c'_1\,,
\label{rhoncft}
\ee
where $a$ is an UV cutoff (e.g. the lattice spacing) and $c_n$ a non-universal constant,
with $c_1=1$.
After the normalization of the two-point function of twist fields on the plane is fixed by the constants $c_n$,
the normalizations of two-point functions in arbitrary geometry and all
other multi-point correlation functions of the twist fields are  fixed and universal. 
Furthermore, although non-universal, the constants $c_n$
satisfy some universal relations (see e.g. \cite{fcm-11}).

\subsection{A single interval.}

Even for conformal invariant theories it is useful to specialize first to the very simple case of a reduced 
density matrix corresponding to a pure system, which is obtained by letting $B\to \emptyset$.
Thus, when $A_2$ is embedded in an infinite system, from Eqs. (\ref{pureqft}) and (\ref{rhoncft})
we have (setting $\ell=u_2-v_2$)
\be\fl
\Tr (\rho_A^{T_2})^{n_e}= (\langle {\cal T}_{n_e/2}(u_2) \overline{\cal T}_{n_e/2}(v_2)\rangle)^2=(\Tr\rho_{A_2}^{n_e/2})^2
=c_{n_e/2}^2 \Big(\frac{\ell}a\Big)^{-{c}/{3}({n_e}/2-2/{n_e})}\,,
\label{1inte}
\ee
and 
\be\fl
\Tr (\rho_A^{T_2})^{n_o}=  \langle {\cal T}_{n_o}(u_2) \overline{\cal T}_{n_o}(v_2)\rangle=\Tr \rho_{A_2}^{n_o}=
c_{n_o} \Big(\frac{\ell}a\Big)^{-{c}/6(n_o-1/{n_o})}.
\label{1into}
\ee
In spite of the simplicity of the above calculation, it shows one important point of the CFT analysis:
for $n=n_e$ even,  
${\cal T}^2_{n_e}$ and $\overline{\cal T}^2_{n_e}$ have dimensions 
\be
\Delta_{{\cal T}_{n_e}^2}=\Delta_{\overline{\cal T}_{n_e}^2}=\frac{c}6 \Big(\frac{n_e}2-\frac2{n_e}\Big), 
\label{Dtne}
\ee
while  for $n=n_o$ odd,  ${\cal T}^2_{n_o}$  and $\overline{\cal T}^2_{n_o}$ have dimensions 
\be 
\Delta_{{\cal T}_{n_o}^2}=\Delta_{\overline{\cal T}_{n_o}^2}=\frac{c}{12}\Big(n_o-\frac1{n_o}\Big),
\label{Dtno}
\ee
the same as ${\cal T}_{n_o}$.  
Thus, performing the analytic continuation from the even branch, we finally have 
\be\fl
|| \rho_A^{T_2}||= \lim_{n_e\to1} \Tr (\rho_A^{T_2})^{n_e}=
c_{1/2}^2  \Big(\frac{\ell}a\Big)^{{c}/2}\qquad  \Rightarrow\;\; {\cal E}=\frac{c}2\ln \frac{\ell}a+2\ln c_{1/2}\,,
\label{neg2pt}
\ee
which, again, just tells us that for pure states the logarithmic negativity equals the 
R\'enyi entropy of order $1/2$.
Continuing instead to $n_o\to1$ from the odd branch we obtain the normalization $\Tr\rho_A^{T_2}=1$.

Notice that although the various constants $c_{n_o,n_e}$ are non-universal, they are the same 
appearing in the entanglement entropies.

\subsection{Two adjacent intervals.}

Let us now consider  the non-trivial configuration in which two intervals $A_1$ and $A_2$ of length $\ell_1$ and $\ell_2$ 
share a common boundary (let us say at the origin) as graphically depicted in Fig. \ref{intervals}.
This can be obtained by letting $v_1\to u_2=0$ in Eq. (\ref{4ptdef}) and it is then described by the  3-point function
(we set $u_1=-\ell_1$ and $v_2=\ell_2$)
\be
\Tr (\rho_A^{T_{2}})^{n}=\langle {\cal T}_{n}(-\ell_1) \overline{\cal T}^2_{n}(0){\cal T}_{n}(\ell_2) \rangle\,,
\ee
whose form is determined by conformal symmetry \cite{cft-book}
\be
\langle {\cal T}_{n}(-\ell_1) \overline{\cal T}^2_{n}(0){\cal T}_{n}(\ell_2) \rangle=
c_n^2 \frac{C_{{\cal T}_{n} \overline{\cal T}^2_{n} {\cal T}_{n}}}{
(\ell_1\ell_2)^{\Delta_{{\cal T}_n^2}} (\ell_1+\ell_2)^{2\Delta_{{\cal T}_n}-\Delta_{{\cal T}_n^2}}},
\label{3ptgen}
\ee
which has been normalized in such a way that the 
structure constant $C_{{\cal T}_{n} \overline{\cal T}^2_{n} {\cal T}_{n}}$ is universal 
(with all the lengths $\ell_j$ measured in units of the UV cutoff $a$)
and can be determined by considering the proper limit of the four-point function,
as it will be done in next section.

For $n=n_e$ even, using the dimensions of the twist operators calculated above, we find
\be
\Tr (\rho_A^{T_{2}})^{n_e}\propto 
{(\ell_1\ell_2)^{-{c}/6({n_e}/2-2/{n_e})} (\ell_1+\ell_2)^{-{c}/6({n_e}/2+1/{n_e})}}\,,
\label{3pteven}
\ee
that in the limit $n_e\to1$ gives
\be
|| \rho_A^{T_{2}}||\propto \left(\frac{\ell_1 \ell_2}{\ell_1+\ell_2}\right)^{{c}/4}\qquad \Rightarrow\quad
{\cal E}= \frac{c}4 \ln \frac{\ell_1\ell_2}{\ell_1+\ell_2}+ {\rm cnst}.
\label{3ptevenbis}
\ee
For $n=n_o$ odd 
\be
\Tr (\rho_A^{T_{2}})^{n_o} \propto (\ell_1\ell_2(\ell_1+\ell_2))^{-\frac{c}{12}(n_o-{1}/{n_o})} ,
\label{3ptodd}
\ee
that for $n_o\to 1$ gives again $\Tr \rho_A^{T_{2}}=1$ as it should. 

Notice that for $n=2$, Eq. (\ref{3pteven}) is 
\be
\Tr (\rho_A^{T_{2}})^2= c_2^2 C_{{\cal T}_{2} \overline{\cal T}^2_{2} {\cal T}_{2}}
 (\ell_1+\ell_2)^{-c/4}\,,
\ee
which equals $\Tr\rho^2_{A_1\cup A_2}$, as predicted by Eq. (\ref{3ptn2}), if 
$ C_{{\cal T}_{2} \overline{\cal T}^2_{2} {\cal T}_{2}}=c_2^{-1}$.

\subsection{Finite systems.}

All the previous results may be generalized to the case of a finite system of length $L$ with periodic boundary conditions
by using a conformal mapping from the cylinder to the plane.
The net effect of the mapping is to replace each length 
 $\ell_i$ with the chord length $(L/\pi) \sin(\pi \ell_i/L)$ in all above formulas.
 
Thus for the case of a pure state in a finite system the generalization of Eqs. (\ref{1inte}) and (\ref{1into}) are
\bea
&&\Tr (\rho_A^{T_2})^{n_e}=
c_{n_e/2}^2 \Big(\frac{L}{\pi a}\sin\frac{\pi \ell}L\Big)^{-{c}/{3}({n_e}/2-2/{n_e})}\,, \label{cftTr1int2}
\\
&&\Tr (\rho_A^{T_2})^{n_o}=
c_{n_o} \Big(\frac{L}{\pi a}\sin\frac{\pi \ell}L\Big)^{-{c}/6(n_o-1/{n_o})}\,, \label{cftTr1int}
\\
&&{\cal E}=\frac{c}2 \ln \Big(\frac{L}{\pi a}\sin\frac{\pi \ell}L\Big)+2\ln c_{1/2}\,.
\label{EN1int}
\eea
For the case of two adjacent intervals the finite system generalizations of Eqs. (\ref{3pteven})  and (\ref{3ptodd}) 
are
\bea\fl
\Tr (\rho_A^{T_{2}})^{n_e}\propto 
\left[\frac{L^2}{\pi^2} \sin \Big(\frac{\pi \ell_1}L \Big) \sin \Big(\frac{\pi \ell_2}L\Big)\right]^{-{c}/6({n_e}/2-2/{n_e})} 
\left[\frac{L}{\pi} \sin \frac{\pi (\ell_1+\ell_2)}L\right]^{-\frac{c}6(\frac{n_e}2+\frac1{n_e})}\,,
\nonumber\\ \fl
\Tr (\rho_A^{T_{2}})^{n_o} \propto \left[
\frac{L^3}{\pi^3} \sin \Big(\frac{\pi \ell_1}L\Big)  \sin \Big(\frac{\pi \ell_2}L\Big) \sin \frac{\pi (\ell_1+\ell_2)}L\right]^{-{c}/{12}(n_o-1/{n_o})} ,
\label{3ptfinL}
\eea
leading to the logarithmic negativity
\be
{\cal E}=\frac{c}4 \ln\left[ \frac{L}{\pi} \frac{\sin \big(\frac{\pi \ell_1}L\big) \sin \big(\frac{\pi \ell_2}L\big)}{\sin\frac{\pi(\ell_1+\ell_2)}L}\right]
+ {\rm cnst}.
\label{neg3finL}
\ee


\section{Negativity for two disjoint intervals in a CFT}
\label{neg2dis}

For the more difficult situation of two disjoint intervals reported in Fig.~\ref{intervals}, 
the entanglement entropies depend on the full operator content of the CFT and this makes much 
more  complicated also the calculation of the negativity.

Global conformal invariance fixes the form of the four-point correlation of twist fields
\be\fl 
 \langle {\cal T}_n(z_1)\overline{\cal T}_n(z_2) {\cal T}_n(z_3)\overline{\cal T}_n(z_4)\rangle=
 c_n^2 \left|\frac{z_{31}z_{42}}{z_{21}z_{43}z_{41}z_{32}} \right|^{{c}/6(n-1/n)} {\cal F}_{n}(x,\bar x)\,,
\label{4ptgen}
\ee
with $x$ the four point ratio
\be
x=\frac{z_{21} z_{43}}{z_{31}z_{42}}\,,
\label{xgen}
\ee
and $z_{ij}=z_i-z_j$. 
Notice that $z_{31} z_{42}/(z_{21} z_{43} z_{41} z_{32}) = 1/(z_{21} z_{43} (1-x))$.
This is normalized in such a way that ${\cal F}_{n}(0,0)=1$ (for $x\to0$, the four-point function is the product
of the two two-point functions normalized with $c_n$). The function ${\cal F}_{n}(x,\bar x)$ depends
explicitly on the full operator content of the theory and must be
calculated case by case. Notice that with ${\cal F}_1(x,\bar x)=1$ as a consequence of $\Tr\rho_A=1$.

Now, according to Eq. (\ref{rhoatw}), $\Tr \rho_A^n$ corresponds in Eq. (\ref{4ptgen}) to the choice
\be
z_1=u_1\,, \quad z_2=v_1\,\quad z_3=u_2\,,\quad z_4=v_2\,,
\ee
and so
\be\fl
\Tr \rho_A^n
=c_n^2 \left(\frac{(u_2-u_1)(v_2-v_1)}{(v_1-u_1)(v_2-u_2)(v_2-u_1)(u_2-v_1)} 
\right)^{{c}/6(n-1/n)} {\cal F}_{n}(x)\,,
\label{Fnd1}
\ee 
where, since for real $u_j$ and $v_j$ the ratio $x$ is real, we dropped the dependence on $\bar x=x$.
Explicitly,  in this case $x$ is  
\begin{equation}
x=\frac{(v_1-u_1)(v_2-u_2)}{(u_2-u_1)(v_2-v_1)}\,,
\label{4pR}
\end{equation}
that for the order $u_1<v_1<u_2<v_2$ satisfies $0<x<1$.

Even $\Tr (\rho_A^{T_{2}})^{n}$ is a four-point function of twist fields according to Eq. (\ref{4ptdef}), but it
corresponds in Eq. (\ref{4ptgen}) to the choice
\be
z_1=u_1\,, \quad z_2=v_1\,\quad z_3=v_2\,,\quad z_4=u_2\,,
\ee
i.e. to the exchange $z_3\leftrightarrow z_4$ compared to $\Tr \rho_A^n$.
Thus we can write
\be\fl
\Tr (\rho_A^{T_2})^n
=c_n^2 \left(\frac{(v_2-u_1)(u_2-v_1)}{(v_1-u_1)(u_2-v_2)(u_2-u_1)(v_2-v_1)} 
\right)^{{c}/6(n-1/n)} {\cal F}_{n}(x)\,,
\label{Fn2d2}
\ee 
where  $x=\frac{(v_1-u_1)(u_2-v_2)}{(v_2-u_1)(u_2-v_1)}$ is again real and now satisfies $-\infty<x<0$. 
This can be rewritten in terms of an ordered four point ratio
\begin{equation}
y=\frac{(v_1-u_1)(v_2-u_2)}{(u_2-u_1)(v_2-v_1)}=\frac{x}{x-1}\,,
\label{4pR2}
\end{equation}
with $0<y<1$, as
\be\fl
\Tr (\rho_A^{T_2})^n
=c_n^2 \left(\frac{(u_2-u_1)(v_2-v_1)}{(v_1-u_1)(v_2-u_2)(v_2-u_1)(u_2-v_1)} 
\right)^{{c}/6(n-1/n)} {\cal G}_{n}(y)\,.
\label{Gn}
\ee 
Equating  equations (\ref{Fn2d2}) and (\ref{Gn}), we can relate the two scaling functions 
${\cal F}_n(x)$ and ${\cal G}_n(y)$ as 
\be
{\cal G}_{n}(y)= (1-y)^{{c}/3\left(n-{1}/n\right)}{\cal F}_{n}\Big(\frac{y}{y-1}\Big)\,.
\label{GvsF}
\ee
Notice that {\it this relation does not depend on the parity of $n$} and so, in order to have a non-trivial
 replica limit $n_e\to1$ for the logarithmic negativity  
\be
{\cal E}(y)=\lim_{n_e\to1} \ln {\cal G}_{n_e}(y)=
\lim_{n_e\to1} \ln \left[ {\cal F}_{n_e}\Big(\frac{y}{y-1}\Big)\right]\,,
\label{NEG}
\ee
some parity effect should appear in the $n$ dependence of the function ${\cal F}_n(x)$ for $x<0$.

The first fundamental consequence of Eq. (\ref{NEG}) is that, 
for conformal invariant systems, the negativity is a scale invariant quantity 
(i.e. a function only of $y$)
because all the dimensional prefactors cancel in the replica limit. 
This has been argued already in the literature on the basis of numerical data \cite{Neg1,Neg2}, but never proved. 

It is useful to consider the ratio
\be
R_n(y)\equiv \frac{\Tr(\rho_A^{T_{2}})^n}{\Tr(\rho_A)^n}=\frac{{\cal G}_n(y)}{{\cal F}_n(y)}=
(1-y)^{c/3(n-1/n)}\frac{{\cal F}_n(y/(y-1))}{{\cal F}_n(y)}
\,,
\label{Rndef}
\ee
in which all the prefactors cancel. 
Being $\displaystyle\lim_{n\to1}\Tr(\rho_A)^n=1$ and ${\cal F}_1(y)=1$ for $0<y<1$, we have that the 
negativity can be obtained just by considering replica limit of this ratio
\be
{\cal E}(y)=\ln \lim_{n_e\to1}  R_{n_e}(y)\,.
\ee

The function ${\cal F}_n(x)$ which gives the negativity has
been calculated for general integral $n$ only for the free compactified boson \cite{cct-09} 
and for the critical Ising model \cite{cct-11}. 
Some other partial results are also known 
\cite{fps-08,cd-08,ch-09,ffip-08,kl-08,atc-10,atc-11,fc-10,c-10,ip-09,h-10,f-12,rg-12,s-12,ms-11,rt-06,hlr-12}. 
However,  the calculations in Refs. \cite{cct-09,cct-11} provide the function ${\cal F}_{n}(x)$ 
only for $0<x<1$ and it is a non-trivial technical problem to extend it to the domain $x<0$ in which we are now interested. 
Before embarking in these calculations, 
we discuss some simple and important physical consequences of Eqs.~(\ref{Gn}), (\ref{GvsF}), and~(\ref{NEG})
which are highlighted by considering the limits $y\to1$ and $y\to0$, i.e.  close and  far intervals respectively.

\subsection{The limit of two very close intervals: $y\to1$.}

If $u_2\to v_1$ then $y\to1^-$ and we should recover the previous result for adjacent intervals.
Let us denote with $\e=u_2-v_1$ and $\ell_j=v_j-u_j$ so that Eq. (\ref{Gn}) becomes
\be\fl
\Tr (\rho_A^{T_2})^n \simeq c_n^2 \left(\frac1{(\ell_1+\ell_2)\e}\right)^{c/6(n-1/n)} {\cal G}_n(y), 
\quad {\rm with}\quad y=1-\frac{\ell_1+\ell_2}{\ell_1 \ell_2} \e+O(\e^2)\,.
\ee 
This reduces to the result for adjacent intervals in Eqs. (\ref{3ptgen}) only if 
\be
{\cal G}_n(y)\simeq g_n (1-y)^{\alpha_n},
\label{Gny1}
\ee
with $\a_n$ equal to 
$\Delta_{{\cal T}_{n}^2}$ the dimension of ${\cal T}_{n}^2$, i.e. 
\be
\a_n=\left\{
\begin{array}{ll}
\displaystyle \frac{c}6\Big(\frac{n_e}2-\frac2{n_e}\Big),& {\rm if}\; n=n_e\; {\rm even},\\
&\\
\displaystyle \frac{c}{12}\Big(n_o-\frac1{n_o}\Big),&  {\rm if}\; n=n_o\; {\rm odd}. 
\end{array}
\right.
\label{alphan}
\ee
We stress that this power-law behavior for $y$ close to $1$ is valid 
apart from possible multiplicative logarithmic corrections, 
whose precise form should be obtained on a case by case basis.

In the replica limit  $n_e\to1$, we then have $\displaystyle \lim_{n_e\to1}\a_{n_e}=-c/4$, 
i.e. the negativity {\it diverges} approaching $y=1$ in the universal way
\be
{\cal E}(y)=-\frac{c}4 \ln (1-y)\,.
\label{Genclose}
\ee 
Clearly for any discrete system with a finite number of degrees of freedom in the two intervals,
the negativity will remain finite in the limit of adjacent intervals, i.e. the limit of zero lattice 
spacing $a\to 0$ and $y\to1$ do not commute (as it is well known when deriving a 
three-point function from the limit of the four-point one \cite{cft-book}).
As it should be clear from the calculation above, this corresponds to set $\epsilon=a$ 
(no distance can be smaller than the lattice spacing) and not true divergence arises 
in a finite discrete system. 
For $n_o\to1$, we instead have $\displaystyle \lim_{n_o\to1}\a_{n_o}=0$ as it should to obtain 
the normalization $\Tr \rho_A^{T_2}=1$. 

Furthermore the knowledge of the function ${\cal G}_n(y)$ allows to calculate 
the universal structure constant $C_{{\cal T}_n \overline{\cal T}_n^2 {\cal T}_n}$ equating Eqs. (\ref{Gny1}) 
and (\ref{3ptgen}), and we obtain (setting $\e=a$)
\be
C_{{\cal T}_n \overline{\cal T}_n^2 {\cal T}_n}=  g_n.
\ee

\subsection{The limit of two very far intervals: $y\to0$.}

The  limit of  far intervals $y\to 0$ can be worked out from the small $x$ expansion of 
${\cal F}_n(x)$ carried out in full generality in Ref.~\cite{cct-11}.
This follows from a generalization of operator product expansion (termed short length
expansion in \cite{cct-11}) that for the case of the entanglement entropies of two intervals give 
\be
{\rm Tr}\,\rho_A^n=
\frac{c_n^2}{(\ell_1\ell_2)^{c/6(n-1/n)}} \sum_{\{k_{j}\}}  \left(\frac{\ell_1\ell_2}{n^2 r^2}\right)^{\sum_j (\Delta_{k_j}+\Db_{k_j})} s_{\{k_j\}}(n)\,,
\label{sle}
\ee
where the sum is over all the sets of possible operators $\{\phi_{k_j}\}$ of the CFT with conformal dimensions 
($\Delta_{k_j}, \Db_{k_j}$). 
$s_{\{k_j\}}(n)$ are calculable coefficients depending on the correlation functions of these operators
\be
s_{\{k_j\}}(n)\propto \langle\prod_{j=1}^n\phi_{k_j}\big(e^{2\pi ij/n}\big)\rangle_{\mathbb C}\,.
\ee 
All these coefficients $s_{\{k_j\}}(n)$, in the limit $n\to1$ (independently of the parity of $n$) 
become one-point functions which always vanish in the complex plane. 
This implies the very strong consequence that the logarithmic negativity
 ${\cal E}(y)$ {\it vanishes as $y\to0$ faster than any power}.

\subsection{The function ${\cal F}_n(x)$ for $0<x<1$.}

In order to calculate $\Tr (\rho_A^{T_2})^n$ we first need to report some known results 
about the function ${\cal F}_n(x)$ which has been calculated for the free compactified boson and
for the Ising model. In the following we will discuss only the case of the boson. 

For a boson compactified on a circle of radius $R$ (a Luttinger liquid field theory), the 
universal scaling function $\mathcal{F}_n(x)$ for generic integral $n\geq1$  and for $0<x<1$
can be written as \cite{cct-09} 
\begin{equation}
\mathcal{F}_n(x)=
\frac{\Theta\big(0|\eta\Gamma\big)\,\Theta\big(0|\Gamma/\eta\big)}{
\Theta\big(0|\Gamma\big)^2}\,,
\label{Fnv}
\end{equation}
where $\Gamma$ is an $(n-1)\times(n-1)$ matrix with elements
\be
\Gamma_{rs} =
\frac{2i}{n} \sum_{k\,=\,1}^{n-1}
\sin\left(\pi\frac{k}{n}\right)\beta_{k/n}\cos\left[2\pi\frac{k}{n}(r-s)\right],
\ee
and
\be
\beta_y=\frac{F_{y}(1-x)}{F_{y}(x)}\,,\qquad
F_{y}(x)\,\equiv\, _2 F_1(y,1-y;1;x)\,,
\ee
$\eta$ is a universal critical exponent proportional to the square of the 
compactification radius $R$, while $\Theta$ is the Riemann-Siegel theta function
\begin{equation}
\label{theta Riemann def}
\Theta(0|\Gamma)\,\equiv\,
\sum_{{\bf m} \in\mathbf{Z}^{n-1}}
\exp\big[i\pi\,{\bf m}\cdot \Gamma \cdot {\bf m}\big]\,.
\end{equation}
This generalizes the result for $n=2$ in Ref. \cite{fps-08}.
We mention that, although the moments of the reduced density matrix have been obtained for all integer $n$, the analytic continuation to
complex $n$ is still beyond our knowledge and so is the Von Neumann entanglement entropy.

It is also worth to report the first terms in the short length expansion (\ref{sle})
in the case of compactified free boson \cite{cct-11}
\be
{\cal F}_n(x)=1+ \left(\frac{x}{4n^2}\right)^{ \a}s_2(n)+ \left(\frac{x}{4n^2}\right)^{2\a}s_4(n)+\dots \,,
\label{Fexpintro}
\ee
where $\a=\min[\eta,1/\eta]$ and 
\be
s_2(n)=n\left(\frac{x}{4n^2}\right)^\a
\sum_{j=1}^{n-1} \frac1{\left[ \sin\left(\pi\frac{j}{n}\right) \right]^{2\a}}\,.
\ee
Also the coefficient $s_4(n)$ has been explicitly calculated \cite{cct-11}.
Both $s_2(n)$ and $s_4(n)$ can be identified with 
the contributions of the short-length expansion coming from the two- and four-point functions 
of the most relevant operators in the theory.  
In the limit $n\to1$, independently of the parity of $n$, we have $s_2(1)=s_4(1)=0$ 
as it should, given  the general considerations above.

\subsection{The function ${\cal G}_n(y)$ and the negativity for a non-compactified boson.}

Before considering the case of the compactified free boson it is worth to study the 
simpler and physical limit when the compactification radius diverges.
In this case, the limit of Eq. (\ref{Fnv}) for $\eta\gg1$ has been worked out 
in Ref.  \cite{cct-09} and it reads for $0<x<1$
\begin{equation}\
{\cal F}_n^{\eta=\infty}(x)= \frac{\eta^{(n-1)/2}}{
\big[\prod_{k=1}^{n-1}F_{k/n}(x)F_{k/n}(1-x)\big]^{1/2}}\,.
\label{decomp} 
\ee 
Despite the relative (compared to Eq.  (\ref{Fnv})) simplicity of this form, 
many physical ingredients for the calculation of the negativity are already present here. 

From Eq. (\ref{GvsF}) we can write ${\cal G}_n(y)$ as
\be\fl
{\cal G}_n^{\eta=\infty}(y)= (1-y)^{(n-1/n)/3}  \frac{\eta^{(n-1)/2}}{
\big[\prod_{k=1}^{n-1}   {\rm Re}\big(F_{k/n}(\frac{y}{y-1}) \overline{F}_{k/n}(\frac{1}{1-y}) \big)\big]^{1/2}},
\label{85}
\ee
and  the ratio with ${\cal F}_n(y)$
\be\fl
R_n^{\eta=\infty}(y)
=(1-y)^{(n-1/n)/3} 
\left[ \frac{
\prod_{k=1}^{n-1}  
F_{{k}/{n}}(y)  F_{{k}/{n}}(1-y)}{\prod_{k=1}^{n-1}  
{\rm Re}\big(F_{{k}/{n}}(\frac{y}{y-1}) \overline{F}_{{k}/n}(\frac{1}{1-y}) \big)}\right]^{1/2},
\label{RnCFT}
\ee
in which the $\eta$ dependence drops out. 
The arguments of the hypergeometric functions in the denominator are not in the 
interval $[0,1]$ and they should be handled with care to select the proper 
analytic continuation valid for $x\in[0,1]$.
This can be achieved by 
 by using that for $x \in (0,1)$ and $k/n \neq 1/2$, it holds
(see e.g. \cite{Bateman} vol. I, section 2.9, Eqs. (34) and (3))
\begin{eqnarray}\fl
F_{k/n}\Big(\frac{1}{1-x}\Big)  &=&
\frac{\Gamma(1-2k/n)}{\Gamma(1-k/n)^2}e^{-i \pi k/n} (1-x)^{k/n}
\, _2F_1\Big(\frac{k}n, \frac{k}n; \frac{2k}n;1-x\Big) + (k \to n-k),
\nonumber \\ \fl
\label{Fkn}
F_{k/n}\Big(\frac{x}{x-1}\Big)  &=&
(1-x)^{k/n} \, _2F_1(k/n, k/n ; 1; x ).
\end{eqnarray}
We stress that these expressions are symmetric under $k \leftrightarrow n-k$.
Instead for $k/n=1/2$ since we have 
\be 
F_{1/2}(z) =  \frac{2}{\pi} K(z), 
\ee
where $K(z)$ is the complete elliptic integral of the first kind, 
we can use the simpler relations
\begin{eqnarray}
\label{F1/2 x/(x-1)}
&& F_{1/2}\Big(\frac{1}{1-x}\Big)  =
\frac{2}{\pi}\sqrt{1-x} \big[ K(1-x)-i K(x) \big],
\\
\label{F1/2 1/(1-x)}
&& F_{1/2}\Big(\frac{x}{x-1}\Big)  =
\frac{2}{\pi}\,\sqrt{1-x} K(x),\\
&& {\rm Re} \Big[ F_{1/2}\Big(\frac{x}{x-1}\Big)\overline{F}_{1/2}\Big(\frac{1}{1-x}\Big)\Big]=
\Big(\frac2\pi\Big)^2 (1-x) K(x) K(1-x)\,. 
\label{FproK}
\end{eqnarray}
From  Eq. (\ref{FproK}) it is clear
that the ratio $R_2^{\eta=\infty}(y)$ equals $1$ identically, as it should according to Eq. (\ref{rhoT2vsrho2}).

For $n>2$, the ratio (\ref{RnCFT}) displays a key mathematical difference between odd and even $n$.
Indeed, each term in the product in both numerator and denominator is invariant under $k \leftrightarrow n-k$ 
(even if this is not evident from the formulas).
Thus, for odd $n$ the products in the numerator and in the denominator are both squares of the products for $k$ up to $(n-1)/2$. 
For even $n$ this is not true because of the presence of the term $k/n=1/2$. 
Nevertheless a simplification occurs between the terms $k/n=1/2$ in the numerator and in the denominator:
\be
\frac{F_{1/2}(y) F_{1/2}(1-y)}{\textrm{Re}\big[F_{1/2}(\frac{y}{y-1}) \overline{F}_{1/2}(\frac{1}{1-y}) \big]}=\frac1{1-y}\,,
\ee
which is equivalent to $R_2(y)=1$.
Thus we have
\be\fl
\label{ratio2}
R_n^{\eta=\infty}(y)
=\left\{
\begin{array}{ll} \displaystyle
(1-y)^{\frac{1}{3}(n-\frac{1}{n})} 
\frac{\prod_{k=1}^{\frac{n-1}{2}}  
F_{k/n}(y) \,F_{k/n}(1-y)}{\prod_{k=1}^{\frac{n-1}{2}}  
\textrm{Re}\big[F_{k/n}(\frac{y}{y-1}) \overline{F}_{k/n}(\frac{1}{1-y}) \big]},
\qquad \textrm{odd $n$},
\\ \\ \displaystyle
(1-y)^{\frac{1}{3}(n-\frac{1}{n})-\frac{1}{2}} 
\frac{\prod_{k=1}^{\frac{n}{2}-1}  
F_{k/n}(y) \,F_{k/n}(1-y)}{\prod_{k=1}^{\frac{n}{2}-1}  
\textrm{Re}\big[F_{k/n}(\frac{y}{y-1}) \overline{F}_{k/n}(\frac{1}{1-y}) \big]},
\qquad \textrm{even $n$},
\end{array}
\right.
\ee

Despite of the relative simplicity of this formula for $R^{\eta=\infty}_n(y)$, 
we did not manage to work out the full analytic continuation to $n_e\to1$. 
However, 
it is extremely instructive to look at the limit $y\to1$ when we can perform
the analytic continuation explicitly and understand how the ``mode'' $k/n=1/2$ is responsible of the 
main qualitative differences between $n_e$ even and $n_o$ odd.

\subsubsection{The analytic continuation in the limit of close intervals.}
In order to calculate the analytic continuation to obtain  the negativity for $y\to 1^-$,  instead of considering
the ratio ${R_{n}^{\eta=\infty}(y)}$, it is easier  to study the function ${\cal G}_n(y)$ in Eq. (\ref{85}) that we rewrite as 
\be
{\cal G}_n(y)= \eta^{(n-1)/2} (1-y)^{1/3(n-1/n)} D_n(y)^{-1},
\label{GvsD}
\ee
with 
\be
D_n(y)=\left[\prod_{k=1}^{n-1}\textrm{Re}\Big[F_{k/n}(\frac{y}{y-1}) \overline{F}_{k/n}(\frac{1}{1-y}) \Big]\right]^{1/2} \,.
\ee
Each term in the product is invariant for $n\leftrightarrow n-k$, thus we can rewrite
it as
\bea\fl \label{Dno}
D_{n_o}(y)&=&\left[\prod_{k=1}^{(n_o-1)/2} {\rm Re}\Big[F_{k/n_o}(\frac{y}{y-1}) \overline{F}_{k/n_o}(\frac{1}{1-y}) \Big]\right] \,,\\ \fl
D_{n_e}(y)&=&\frac2\pi (1-y)^{1/2}
\sqrt{K(y)K(1-y)}\left[\prod_{k=1}^{n_e/2-1} {\rm Re}\Big[F_{k/n_e}(\frac{y}{y-1}) \overline{F}_{k/n_e}(\frac{1}{1-y}) \Big]\right] ,
\label{Dne}
\eea
where for $n=n_e$ even we isolated the term $k/n=1/2$ using Eq. (\ref{FproK}).

In order to calculate the asymptotic behavior as $y\to{1^-}$, we rewrite  each factor in the two products above as
\begin{eqnarray}\label{Fexp}
\fl
{\rm Re}\big[F_{k/n}&& (\frac{y}{y-1})  \overline{F}_{k/n}(\frac{1}{1-y}) \big]
= (1-y)^{k/n} \, _2F_1(k/n, k/n ; 1; y )
\\ \fl
& & 
\times \Big(
\frac{\Gamma(1-{2k}/{n})}{\Gamma(1-{k}/{n})^2} (1-y)^{{k}/{n}}
\, _2F_1(\frac{k}{n}, \frac{k}{n}; \frac{2k}{n};1-y) - (k \leftrightarrow n-k)
\Big) \cos(\pi k/n).
\nonumber
\end{eqnarray} 
Then, for the asymptotic behavior of $D_n(y)$ close to $y=1$ we need 
\bea
\fl&&
\lim_{x\to1^-}\, _2F_1({k}/{n}, {k}/{n};1;x)=
\frac{\pi}{\sin(2\pi \frac{k}{n})\, \Gamma(1-\frac{k}{n})^2\, \Gamma(\frac{2k}{n})},
\qquad
\frac{k}{n}<\frac12\,,\\ \fl&&
\lim_{x\to1^-} K(x) =-\frac{1}{2}\ln(1-x) + 2\ln 2+ o(1),
\label{Kexp}\\\fl &&
K(0)=\frac\pi2,\qquad \, _2F_1(a,b;c;0)=1.
\eea
For $(1-y)\ll1$, in Eq. (\ref{Fexp}) the term $(1-y)^{k/n}$ dominates compared to $(1-y)^{1-k/n}$ 
(we recall $k/n<1/2$) which can be neglected.

Thus for  odd $n$, the  expansion of (\ref{Dno}) as $y \rightarrow 1$ is 
\begin{equation}\fl 
D_{n_o}(y)
=(1-y)^{ \left(n_o-1/n_o\right)/4}
 \prod_{k=1}^{(n_o-1)/2}
\frac{\pi \Gamma(1-\frac{2k}{n_o})}{2 \sin(\pi \frac{k}{n_o}) \Gamma(1-\frac{k}{n_o})^4 \Gamma(\frac{2k}{n_o})}
 +  \dots,
\end{equation}
where we used Eq. (\ref{Fexp}) and $\prod_{k=1}^{(n-1)/2} x^{2k/n}=x^{(n-1/n)/4}$.
After some algebra, we have $\displaystyle \lim_{n_o\to1}D_{n_o}(y)=1$, as it should. 
Plugging the value of $D_{n_o}(y)$ into Eq. (\ref{GvsD}), we recover the general expansion for $y$
close to $1$ in Eqs. (\ref{Gny1}) and (\ref{alphan}).

For even $n$, using again Eq. (\ref{Fexp}) and $\prod_{k=1}^{n/2-1} x^{2k/n}=x^{n/4-1/2}$,
 we  have 
\be
D_{n_e}(y)= 
(1-y)^{n_e/4} \sqrt{ K(y)} P_{n_e}+\dots \,,
\ee
where we defined the constant 
\be
P_n = \sqrt{\frac2\pi} \prod_{k=1}^{n/2-1}
\frac{\pi \Gamma(1-\frac{2k}{n})}{2 \sin(\pi \frac{k}{n}) \Gamma(1-\frac{k}{n})^4 \Gamma(\frac{2k}{n})},
\ee
and  we left $K(y)$ instead of considering its expansion in 
Eq. (\ref{Kexp}),  for simplicity in the notations.
Plugging  $D_{n_e}(y)$ in Eq. (\ref{GvsD}), we recover the general expansion for $y$ close 
to $1$ of ${\cal G}_{n_e}(y)\propto (1-y)^{c/6(n_e/2-2/n_e)}$ up to a logarithmic correction. 

For $n_e\to 1$, we then have
\be
 D_1(y) =  (1-y)^{1/4} \sqrt{ K(y)} P_1 + o(1),
\ee 
which gives the desired expansion of the logarithmic negativity for $y\to1^{-}$
\bea\fl
\label{Eclose}
{\cal E}(y)&=&-\ln D_1(y)= 
-\frac14 \ln (1-y) - \frac12 \ln K(y) -\ln  P_1 +o(1)\\
\fl &=& -\frac14 \ln (1-y) - \frac12 \ln \Big(-\frac12 \ln (1-y)\Big) -\ln  P_1 +o(1).\nonumber
\eea 
(Notice a typo in the published version of Ref. \cite{us-letter} where we wrote the expansion of $e^{{\cal E}(y)}$ instead 
of ${\cal E}(y)$). 
A part from a subleading double-logarithmic correction, this agrees with the general expansion 
given in Eq. (\ref{Genclose}) for the negativity of two intervals when they get closer. 
Here we can also calculate analytically the constant $P_1$. 

From the definition of $P_n$  we have
\be\fl
\sqrt{\frac\pi2}P_n\equiv  \prod_{k=1}^{n/2-1}\frac{\pi \Gamma(1-\frac{2k}{n})}{2 \sin(\pi \frac{k}{n}) \Gamma(1-\frac{k}{n})^4 \Gamma(\frac{2k}{n})}=
 \prod_{k=1}^{n/2-1}\frac{\pi}{2 \sin(\pi \frac{k}{n}) \Gamma(1-\frac{k}{n})^4}\,,
 \label{Pn1}
\ee
where we used $ \prod_{k=1}^{n/2-1}  [\Gamma(1-{2k}/{n})/  \Gamma({2k}/{n})]=1$ 
because in the product the same terms appear in opposite order in the numerator and in the denominator. 
Part of the remaining product is trivial and indeed we have
\be
 \sqrt{\frac2\pi}  \prod_{k=1}^{n/2-1}\frac{\pi}{2 \sin(\pi k/n)}=  \Big(\frac{\pi}2\Big)^{(n-3)/2}\frac1{n^{1/2} 2^{(1-n)/2}}
 \stackrel{n\to 1}{\longrightarrow}\frac2\pi \,. 
 \label{1cont}
\ee 
The second piece is instead more complicated.
Let us consider the product 
\be
p_n=\prod_{k=1}^{n/2-1} \Gamma(1-k/n)\Rightarrow \ln p_n= \sum_{k=1}^{n/2-1} \ln \Gamma(1-k/n)\,.
\ee
Now we can employ the following integral representation of the logarithm of the $\Gamma$ function 
 \be
\ln \Gamma(z) =\int_0^\infty\frac{dt e^{-t}}{t}\left( 
\frac{e^{-(z-1) t} - 1}{1 - e^{-t}} + z - 1\right)\,,
\ee
 to rewrite
\be
\ln p_n=  \sum_{k=1}^{n/2-1} \int_0^\infty \frac{dt e^{-t}}{t}\left( \frac{e^{tk/n} - 1}{1 - e^{-t}}-\frac{k}n\right),
\ee
which, inverting the order of sum and integral, can be easily summed up as
\be
\ln p_n =\int_0^\infty \frac{dt e^{-t}}{t} \left[\frac1{1 - e^{-t}} \left(\frac{e^{t/2}-1}{  e^{t/n}-1}-\frac{n}2\right)-\frac{n-2}8\right]\,, 
\ee
whose analytic continuation at $n=1$ is 
\be\fl
 p_1 =\exp\left(\int_0^\infty \frac{dt e^{-t}}{t} \left[\frac1{1 - e^{-t}} \left(\frac{e^{t/2}-1}{  e^{t}-1}-\frac{1}2\right)+\frac18\right]\right)=
\frac{A^{3/2}}{{2}^{1/24} e^{1/8} {\pi^{1/4} }},
\label{p1}
\ee
where $A= \exp(1/12- \zeta'(-1)) = 1.2824 \dots$  is GlaisherÕs constant, related to the Riemann zeta function $\zeta(z)$.
 Putting together Eqs. (\ref{Pn1}), (\ref{1cont}),  and (\ref{p1}) we finally have
\be
  P_1=\frac2{\pi p_1^4}=\frac{2^{7/6} e^{1/2}}{A^6}= 0.832056\dots\,.
  \label{PP1}
\ee

\subsection{The function ${\cal G}_n(y)$ and the negativity for the compactified boson.}

As discussed in  Ref. \cite{cct-09}, for a free complex boson whose real and imaginary parts are compactified 
on a circle of radius $R$ when encircling a branch point, we have the additional freedom of winding around 
the circle in the target space, i.e. $\phi_j(e^{2\pi i} z, e^{-2\pi i} \bar{z}) =\phi_{j-1}(z, \bar{z}) + 2\pi R(m_{j,1}+i m_{j,2})$, 
where $m_{j,1} , m_{j,2} \in \mathbb{Z}$, if the branch point is in the origin.
The integer numbers $m_{i,l}$ can be organized into vectors such as ${\bf m}_1=(m_{1,1},m_{2,1}\dots m_{n,1})$
or even in $2n$-dimensional ones ${\bf m}=(m_{1,1}, m_{2,1}\dots m_{n,1},m_{1,2}, m_{2,2}\dots m_{n,2})$.
In all this section we will denote vectors with bold symbols to be easily distinguished from scalars.

The function ${\cal F}_n(x,\bar x)$ appearing in Eq.  (\ref{4ptgen})  for a real scalar field is the square root of the 
same quantity for a complex field because the latter is the sum of two independent real scalar fields. 
Then, following  \cite{cct-09} and using some results in \cite{Dixon:1986qv,z-87}, we have that the square of 
${\cal F}_n(x,\bar x)$ for a real scalar field and for any four-point ratio $x, \bar x \in \mathbb{C}$ is given by 
\begin{equation}\fl
\label{Fn sum}
[{\cal F}_n(x,\bar x)]^2 =
\prod_{k=1}^n  
\frac{\rm const}{F_{k/n}(x) \bar{F}_{k/n}(1-\bar{x}) + \bar{F}_{k/n}(\bar{x}) F_{k/n}(1-x)}
 \sum_{{\bf m}^{(1)},{\bf m}^{(2)} } \prod_{k=1}^n  e^{- S_{k/n}^{\rm cl}},
\end{equation}
where the sum runs over the possible integer components of 
${\bf m}^{(1)},{\bf m}^{(2)} \in ({\mathbb Z+i \mathbb Z})^{n}={\mathbb Z}^{2n}$
and where the constant does not depend on $x$ and it will be fixed later on.
The dependence on the compactification radius is encoded in the ``classical action'' 
$S_{k/n}^{\rm cl}$ first calculated  in Ref. \cite{Dixon:1986qv}:
\begin{equation}\fl
\label{classical action}
 S_{k/n}^{\rm cl} =
 \frac{2g \pi \sin(\pi k/n)}{n}
\left[
\frac{|\tau_{k/n}|^2}{\beta_{k/n}}  |\xi_1|^2
+\frac{\alpha_{k/n}}{\beta_{k/n}} (\xi_1\bar{\xi}_2 \bar{\gamma}+\bar{\xi}_1\xi_2 \gamma)
+ \frac{|\xi_2|^2}{\beta_{k/n}} \right] \,,
\end{equation}
where $g$ is  the Lagrangian coupling $\mathcal{L}[\phi_j] =g |\nabla \phi_j|^2/(4\pi)$ and $\gamma \equiv - i e^{-i \pi k/n}$
and  $\xi_p$ ($p=1,2$) are 
\begin{equation}
\label{xi vectors}
\xi_p = 2\pi  R \sum_{l=1}^{n} (\theta_{k/n})^l
\big(m_{l,1}^{(p)}+ i\, m_{l,2}^{(p)}\big),
\qquad
 \theta_{k/n} \equiv e^{2\pi i k/n}\,.
\end{equation}
The dependence on the harmonic ratio $x$ in Eq. (\ref{xgen}) is instead encoded in the 
quantity $\tau_{k/n}$ also derived in Ref. \cite{Dixon:1986qv}: 
\begin{equation}
\tau_{k/n}=i \frac{F_{k/n}(1-x)}{F_{k/n}(x)}
\equiv  \alpha_{k/n}+ i \beta_{k/n}\,,
\end{equation}
being $\alpha_{k/n}$ and $\beta_{k/n}$ the real and the imaginary part of $\tau_{k/n}$ respectively.
Before embarking in the main calculation, we stress that the crucial difference compared to the 
case of real $x$ with the constraint $0\leq x\leq 1$, considered in Ref. \cite{cct-09}, 
is the presence of the linear term in $\gamma$ and $\bar \gamma$
in Eq. (\ref{classical action}) which was vanishing because $\alpha_{k/n}=0$ for $0\leq x \leq 1$.

The products involving $\xi_p$ in Eq. (\ref{classical action}) can be written as 
\begin{eqnarray}\fl
\label{xip xiq}
\frac{\xi_p \bar{\xi}_q}{(2\pi R)^2} & = & 
 \sum_{r,s = 1}^n\hspace{-1mm}
(\tilde{C}_{k/n})_{rs}
[ m_{r,1}^{(p)} m_{s,1}^{(q)}+ m_{r,2}^{(p)} m_{s,2}^{(q)}] 
-  (\tilde{S}_{k/n})_{rs}
[ m_{r,2}^{(p)} m_{s,1}^{(q)}- m_{r,1}^{(p)} m_{s,2}^{(q)}]
\\ \fl
& & \hspace{0.45cm}+ i \left[
 (\tilde{S}_{k/n})_{rs}
[ m_{r,1}^{(p)} m_{s,1}^{(q)}+ m_{r,2}^{(p)} m_{s,2}^{(q)}]
+ (\tilde{C}_{k/n})_{rs}
[ m_{r,2}^{(p)} m_{s,1}^{(q)} - m_{r,1}^{(p)} m_{s,2}^{(q)}]
\right], 
\nonumber
\end{eqnarray}
where we introduced the $n \times n$ real matrices $\tilde{C}_{k/n}$ and $\tilde{S}_{k/n}$ whose elements are
\begin{equation}\fl 
(\tilde{C}_{k/n})_{rs} \equiv  \cos\Big(2\pi \frac{k}{n}(r-s)\Big)\,,
\qquad
(\tilde{S}_{k/n})_{rs} \equiv  \sin\Big(2\pi \frac{k}{n}(r-s)\Big)\,,
\ee
with $r,s=1, \dots , n$.
Notice that $\tilde C_{k/n}$ is  symmetric and $\tilde S_{k/n}$ is antisymmetric.
Plugging (\ref{xip xiq}) into (\ref{classical action}) and summing over $k$, as required in (\ref{Fn sum}), we obtain
\begin{eqnarray}\fl
\sum_{k = 1}^n S_{k/n}^{\rm cl} & = & 
\frac{8g R^2 \pi^3}{n} \sum_{r,s =1}^n \sum_{k=1}^n  \sin(\pi \frac{k}{n})
 \Bigg\{ 
\frac{|\tau_{k/n}|^2}{\beta_{k/n}}\, \big(\tilde{C}_{k/n}\big)_{rs}
\left( m_{r,1}^{(1)} m_{s,1}^{(1)}+ m_{r,2}^{(1)} m_{s,2}^{(1)}\right)  \nonumber \\ \fl
& & 
- \frac{2\alpha_{k/n}}{\beta_{k/n}}\bigg[
\cos(\pi \frac{k}{n}) \big(\tilde{S}_{k/n}\big)_{rs} \left( m_{r,1}^{(1)} m_{s,1}^{(2)} + m_{r,2}^{(1)} m_{s,2}^{(2)}\right)
 \nonumber  \\ \fl && \qquad 
+\sin(\pi \frac{k}{n}) \big(\tilde{C}_{k/n}\big)_{rs} \left( m_{r,1}^{(1)} m_{s,1}^{(2)} + m_{r,2}^{(1)} m_{s,2}^{(2)}\right)
\bigg] \nonumber \\ \fl
& & 
+ \frac{1}{\beta_{k/n}}\, \big(\tilde{C}_{k/n}\big)_{rs}
\left( m_{r,1}^{(2)} m_{s,1}^{(2)}+ m_{r,2}^{(2)} m_{s,2}^{(2)}\right)
\Bigg\}\,, \label{sum Scl}
\end{eqnarray}
where many terms of $S_{k/n}^{\rm cl}$ vanish in the sum over $k$ because they are odd under $k \leftrightarrow n-k$.
Introducing the $2n$ vectors of integers ${ \tilde{\bf m}}_1 \equiv ({\bf m}_1^{(1)} , {\bf  m}_1^{(2)})$ 
and ${ \tilde{\bf m}}_2 \equiv ({\bf m}_2^{(1) } , {\bf m}_2^{(2)})$ 
made respectively by the real and the imaginary parts of ${\bf m}^{(1)}$ and ${\bf m}^{(2)}$, it is possible
to rewrite Eq. (\ref{sum Scl}) as
\begin{equation}\fl
\label{Scl Gtilde}
- \sum_{k=1}^n S_{k/n}^{\rm cl} 
= i \pi \eta
\Big(
{ \tilde{\bf m}}_1^{\rm t}  \cdot \tilde{G} \cdot { \tilde{\bf m}}_1 
+
{ \tilde{\bf m}}_2^{\rm t}  \cdot \tilde{G} \cdot { \tilde{\bf m}}_2 
\Big),
\qquad
{ \tilde{\bf m}}_1   , { \tilde{\bf m}}_2 \in \mathbb{Z}^{2n},
\end{equation}
where $\eta \propto  R^2$. 
The matrix $\tilde{G} $ is symmetric, purely imaginary and it can be written in terms of $n \times n$ block matrices as
\begin{equation}
\label{Gt def}
\tilde{G} \equiv 2 i 
\left(\begin{array}{cc}
\tilde{A} & \tilde{W} \\ 
\tilde{W}^{\rm t} & \tilde{B}
\end{array}\right)\,,
\end{equation}
with
\begin{eqnarray}
\label{At def}\fl
\tilde{A} &= & \frac{1}{n}
\sum_{k=1}^n 
\frac{|\tau_{k/n}|^2}{\beta_{k/n}}\sin(\pi k/n) \, \tilde{C}_{k/n}
=
\sum_{k=1}^n 
\frac{|\tau_{k/n}|^2}{\beta_{k/n}}\sin(\pi k/n) \, \tilde{E}_{k/n}\,,
\\ \fl 
\label{Bt def}
\tilde{B} &= & \frac{1}{n}
\sum_{k=1}^n 
\frac{1}{\beta_{k/n}}\sin(\pi k/n) \, \tilde{C}_{k/n}
=
\sum_{k=1}^n 
\frac{1}{\beta_{k/n}}\sin(\pi k/n) \, \tilde{E}_{k/n}\,,
\\ \fl 
\label{Wt def}
\tilde{W} &= & 
- \,\frac{1}{n} \sum_{k=1}^n 
\frac{\alpha_{k/n}}{\beta_{k/n}} \sin(\pi k/n) 
\Big[ \sin(\pi k/n) \, \tilde{C}_{k/n} + \cos(\pi k/n)  \tilde{S}_{k/n}  \Big]
\\ \fl 
& = &
- \sum_{k=1}^n 
\frac{\alpha_{k/n}}{\beta_{k/n}} \sin(\pi k/n) 
\big[  \sin(\pi k/n)  -i  \cos(\pi k/n)   \big] \tilde{E}_{k/n}\,,
\nonumber
\end{eqnarray}
where we introduced the complex matrices $\tilde{E}_{k/n}$ for $k=1,\dots , n$ whose elements are 
\be (\tilde{E}_{k/n})_{rs} \equiv e^{2\pi i (k/n)(r-s)}/n.
\ee 
These are hermitian matrices satisfying $\tilde{E}_{p/n} \cdot \tilde{E}_{q/n} = \delta_{p,q} \tilde{E}_{p/n}$, therefore they are projectors,
a property which guarantees that $\tilde{A}$, $\tilde{B}$ and $\tilde{W}$ commute.
Since $\tilde{A}$ and $\tilde{B}$ are symmetric, $\tilde{G}$ is symmetric by construction. 
The eigenvalues of $\tilde{A}$, $\tilde{B}$ and $\tilde{W}$ are respectively
\begin{equation}\fl 
\label{eigenvalues abg}
a_{k} = \frac{|\tau_{k/n}|^2}{\beta_{k/n}}\sin(\pi \frac{k}n) ,
\qquad
b_{k}= \frac{1}{\beta_{k/n}}\sin(\pi \frac{k}n) ,
\qquad
w_{k} = i\frac{\alpha_{k/n}}{\beta_{k/n}} \sin(\pi \frac{k}n) e^{i \pi  k/n},
\end{equation}
where $k = 1, \dots , n$. 
Most of these eigenvalues  are degenerate because of the symmetry $k \leftrightarrow n-k$. 
Since $\tilde{A}$, $\tilde{B}$ and $\tilde{W}$ commute, the characteristic polynomial of $\tilde{G}$ is
\begin{eqnarray}\fl \label{characteristic pol step3}
P(\lambda) &=& 
\textrm{det}\Big[(\tilde{A}- \lambda {\mathbb I}_n)(\tilde{B}- \lambda {\mathbb I}_n)
- \tilde{W} \cdot \tilde{W}^{\rm t} \Big] 
\\ \fl 
&=&
\textrm{det}\left(
\sum_{k=1}^n 
\Big[ (a_{k} -\lambda)(b_{k} -\lambda)-|w_{k} |^2 \Big] \tilde{E}_{k/n}
\right)
=\prod_{k=1}^n 
\Big[ (a_{k} -\lambda)(b_{k} -\lambda)-|w_{k} |^2 \Big] , \nonumber
\end{eqnarray}
which is obtained from the fact that the matrices $\tilde{E}_{k/n}$ are projectors.
Then the eigenvalues of $\tilde{G}$ are
\begin{eqnarray}
\label{eigenvalues wt}
\lambda_{k}^{\pm} & =& \frac{1}{2}
\left(a_{k} + b_{k} \pm \sqrt{(a_{k} - b_{k})^2+4|w_{k} |^2} \right)
\\
\label{lambdapm}
&=&
\frac{\sin(\pi k/n)}{2 \beta_{k/n}}
\left( |\tau_{k/n}|^2+1\pm \sqrt{ (|\tau_{k/n}|^2+1)^2-4 \beta_{k/n}^2} \right)\,.
\end{eqnarray}
They are all real and strictly positive for $k \in \{1, \dots , n-1\}$, while $\lambda_n^{\pm}  =0 $. 
We also have $\lambda_k^{\pm}  =\lambda_{n-k}^{\pm}  $ for the invariance $k \leftrightarrow n-k$ of (\ref{eigenvalues abg}).
Thus, 0 is always a doubly degenerate eigenvalue. As for the degeneracy of the remaining positive eigenvalues, if $n-1$ is even then they are all doubly degenerate as well; instead if $n-1$ is odd then they are all doubly degenerate except for $\lambda^\pm_{n/2}$ which are non degenerate.

From the common eigenvectors of $\tilde{A}$, $\tilde{B}$ and $\tilde{W}$, we can construct the $2n \times 2n $ matrix $\tilde{U}$ diagonalizing $\tilde{G}$ 
\begin{equation}\fl
\label{Gt diag}
\tilde{G} \,=\,\tilde{U} \cdot
\left(\begin{array}{cccc|cccc}
\lambda_{1}^{+} & & & & & & &  \\
 & \ddots & & & & & &  \\
 & & \lambda_{n-1}^{+}  & & & & &  \\ 
  & &  & 0 & & & &  \\ 
 \hline
  & & & & \lambda^{-}_{1} & & &  \\ 
  & & & &  & \ddots & &  \\ 
  & & & & & &  \lambda^{-}_{n-1} &  \\ 
    & & & &  & & & 0  \\ 
\end{array}\right)\cdot \, \tilde{U}^{-1}  \,.
\end{equation}
Since $\tilde{G}$ has zero eigenvalues, 
the sum over the vectors ${\bf m}^{(j)}$ in Eq. (\ref{Fn sum}) is divergent. 
Thus, as done in \cite{cct-09}, we introduce a regularized $\tilde{G}_\epsilon$ by replacing the two zeros on the diagonal of the diagonal matrix in (\ref{Gt diag}) with a small cutoff $\epsilon >0$.
This allows us to write the regularized sum in (\ref{Fn sum}) in terms of a Riemann Siegel theta function 
\begin{equation}\fl 
\sum_{{ \tilde{\bf m}}_1   ,  \tilde{\bf m}_2 \, \in\,  \mathbb{Z}^{2n}} 
\exp \left[  i \pi \eta
\Big(
{ \tilde{\bf m}}_1^{\rm t}  \cdot \tilde{G}_\epsilon \cdot {\tilde{\bf m}}_1  +
{ \tilde{\bf m}}_2^{\rm t}  \cdot \tilde{G}_\epsilon  \cdot {\tilde{\bf m}}_2 
\Big)
\right]
= \Theta\big(0| \eta \tilde{G}_\epsilon \big)^2\,,
\end{equation}
where the regularized $\tilde{G}_\epsilon$ can be written as
\begin{equation}
\tilde{G}_\epsilon = \tilde{G}+\frac{\epsilon}{n} 
\left(\begin{array}{cc}
{I}_n & 0_n \\
0_n & {I}_n 
\end{array}\right)\,,
\end{equation}
where ${I}_n$ and   $0_n$ are $n \times n$ matrices with all equal elements given by  1 and 0 respectively.
Now we can isolate the divergent factor by observing that the  limit 
\begin{equation}
\label{lim theta}
\lim_{\epsilon \to0}\big[ n \epsilon \Theta\big(0| \eta \tilde{G}_\epsilon \big)\big]=
\Theta\big(0| \eta G \big)\,,
\end{equation}
gives a finite result.
Since the divergent part is independent of $x$, it can be adsorbed into a normalization constant.
The $2(n-1) \times 2(n-1)$ symmetric matrix $G$ introduced in the r.h.s. of (\ref{lim theta})  is
\begin{equation}
\label{G def}
G  \equiv 2 i
\left(\begin{array}{cc}
A & W \\ 
W^{\rm t} & B
\end{array}\right)\,,
\end{equation}
where
\begin{eqnarray}\fl 
\label{A def}
A &= & \frac{1}{n}
\sum_{k = 1}^{n-1}
\frac{|\tau_{k/n}|^2}{\beta_{k/n}}\sin(\pi k/n) \, C_{k/n}
=
\sum_{k=1}^{n-1}
\frac{|\tau_{k/n}|^2}{\beta_{k/n}}\sin(\pi k/n) \, E_{k/n}\,,
\\ \fl 
\label{B def}
B &= & \frac{1}{n}
\sum_{k = 1}^{n-1}
\frac{1}{\beta_{k/n}}\sin(\pi k/n) \, C_{k/n}
=
\sum_{k\,=\,1}^{n-1} 
\frac{1}{\beta_{k/n}}\sin(\pi k/n) \, E_{k/n}\,,
\\ \fl 
\label{W def}
W &= & 
- \frac{1}{n} \sum_{k = 1}^{n-1} 
\frac{\alpha_{k/n}}{\beta_{k/n}} \sin(\pi k/n)  \big[ \sin(\pi k/n)  C_{k/n} + \cos(\pi k/n)  S_{k/n} \big]
\\ \fl
& = &
- \sum_{k = 1}^{n-1} 
\frac{\alpha_{k/n}}{\beta_{k/n}} \sin(\pi k/n) 
\big[  \sin(\pi k/n)  -i  \cos(\pi k/n)   \big] E_{k/n}\,.
\nonumber
\end{eqnarray}
These matrices are obtained from the corresponding tilded ones given in Eqs. (\ref{At def}), (\ref{Bt def}) and (\ref{Wt def}) 
by removing the last column and the last row, namely
\begin{equation}\fl 
\big(C_{k/n}\big)_{rs} \equiv  \big(\tilde{C}_{k/n}\big)_{rs}\,,
\hspace{.5cm}
\big(S_{k/n}\big)_{rs} \equiv  \big(\tilde{S}_{k/n}\big)_{rs}\,,
\hspace{.5cm}
\big(E_{k/n}\big)_{rs} \equiv  \big(\tilde{E}_{k/n}\big)_{rs}\,,
\ee
with $r,s=1, \dots , n-1$.
Absorbing the divergent part into the normalization constant, we can finally write (\ref{Fn sum}) as
\begin{equation}
\label{Fn}\fl 
[{\cal F}_n(x, \bar x)]^2
=
\textrm{const}
\frac{\Theta(0 | \eta G )^2}{\prod_{k=1}^{n-1}  \big[F_{k/n}(x) \,\bar{F}_{k/n}(1-\bar{x}) + \bar{F}_{k/n}(\bar{x}) \,F_{k/n}(1-x)\big]}\,.
\end{equation}
As first check,  for real $x$ we have $\alpha_{k/n}=0$ for every $k$ and therefore the off diagonal blocks $W$ in $G$ are zero. 
In this case the Riemann theta function $\Theta(0 | \eta G)$ factorizes into a product of two Riemann theta functions  
and (\ref{Fn}) reproduces the result of \cite{cct-09}, as it should. 

As for the normalization constant in (\ref{Fn}), since it is independent of $x$, it is given by the one computed in \cite{cct-09}, 
namely $(2\eta)^{n-1}$. Thus, for the real compactified free boson, we find that
\begin{equation}
\label{Fn1}\fl 
{\cal F}_n(x,\bar x)=
\frac{\eta^{(n-1)/2} \Theta(0 | \eta G )}{
\sqrt{ \prod_{k=1}^{n-1}  {\rm Re}(F_{k/n}(x)  \bar{F}_{k/n}(1-\bar{x}) )}}
=
\frac{\eta^{(n-1)/2} \Theta(0 | \eta G )}{\prod_{k=1}^{n-1}  
\big|F_{k/n}(x) \big| \sqrt{{\rm Im}(\tau_{k/n}(x))}},
\end{equation}
where we used that ${\rm Re}(F_{k/n}(x) \bar{F}_{k/n}(1-\bar{x})) = |F_{k/n}(x) |^2  {\rm Im}(\tau_{k/n}(x))$. 
In the limit of non compactified boson $\eta \to \infty$ the Riemann-Siegel theta function does not contribute because 
$\Theta(0 | \eta G ) \to 1$,  recovering the results for the non-compactified boson. 
Notice also that for $\eta=1$, while for $x$ real and $x\in [0,1]$ we have ${\cal F}_n(x,\bar x)=1$ \cite{cct-09}, this is 
not true for general complex $x$ and, in particular, for $x$ real and negative.

From Eqs. (\ref{GvsF}) and (\ref{Fn1}) we can write $\mathcal{G}_n(y)$ for $0<y<1$ as
\begin{equation}\fl 
\label{Gn compact}
\mathcal{G}_n(y)
= (1-y)^{(n-1/n)/3}
\frac{\eta^{\frac{n-1}{2}}}{\big[ \prod_{k=1}^{n-1}  
\textrm{Re}\big(F_{k/n}(\frac{y}{y-1})   \bar{F}_{k/n}(\frac{1}{1-y}) \big)\big]^{1/2}}
 \Theta\Big(0 \big| \eta G \big(\frac{y}{y-1}\big)\Big).
\end{equation}
Unfortunately, we are not able to compute the analytic continuation of ${\cal G}_{n_e}(y)$ to $n_e \to 1$ which gives the logarithmic negativity for the free compactified boson.
This reflects a similar problem for the standard R\'enyi entropies which are also known only for 
integer $n>2$.

\subsubsection{Invariance of ${\cal F}_n(x,\bar x)$ for $x \leftrightarrow 1-x$ and for $\eta \leftrightarrow 1/\eta$.}

For real $x$, it turned out that ${\cal F}_n(x)$ was invariant under $x \leftrightarrow 1-x$.
In order to show this invariance also for complex $x$, we focus on Eq. (\ref{Fn}) in which  
the denominator is clearly invariant. 
As for its numerator $\Theta(0 | \eta G )$, it is also invariant because when $x \leftrightarrow 1-x$ we have
\begin{equation} \fl 
\tau_{k/n} \to -\frac{1}{\tau_{k/n}}
\hspace{.7cm}
\Longleftrightarrow
\hspace{.7cm}
\alpha_{k/n} \to -\frac{\alpha_{k/n} }{|\tau_{k/n}|^2},
\hspace{.8cm}
\beta_{k/n} \to \frac{\beta_{k/n} }{|\tau_{k/n}|^2}\,,
\end{equation}
leading to 
\begin{equation}
A \,\rightarrow\,B\,,
\hspace{1.5cm}
W \,\rightarrow\,-W\,.
\end{equation}
This change does not modify the sum defining $\Theta(0 | \eta G )$. Indeed, it can be reabsorbed by a redefinition of the sign of the first half (or, equivalently, of the second half) of the vector of integers over which we are summing.

The function $\mathcal{F}_n(x,\bar x)$ in (\ref{Fn1}) is also invariant under $\eta \leftrightarrow 1/\eta$.
Indeed, by employing the Poisson summation formula, we can rewrite (\ref{Fn1}) in a form where this invariance is manifest, namely
\begin{equation}
\label{Fn2}
\mathcal{F}_n(x,\bar x) =
\frac{\Theta(0 | T)}{\prod_{k=1}^{n-1}  
|F_{k/n}(x) |}\,,
\end{equation}
where the $2(n-1)\times 2(n-1)$ symmetric matrix $T$ is 
\begin{equation}
\label{matrix T}
T =
\left(\begin{array}{cc}
i \eta \mathcal{I} & \mathcal{R}  \\ 
\mathcal{R} & i \mathcal{I} / \eta
\end{array}\right)\,,
\end{equation}
and the $(n-1)\times (n-1)$ symmetric matrices $ \mathcal{I}$ and $ \mathcal{R}$ are
\begin{eqnarray}
\label{I matrix def}
 \mathcal{I} &= & \frac{2}{n}
\sum_{k=1}^{n-1}
\beta_{k/n} \sin(\pi k/n) \, C_{k/n}\,
=
 2\sum_{k=1}^{n-1}
\beta_{k/n}  \sin(\pi k/n) \, E_{k/n}\,,
\\
\label{R matrix def}
 \mathcal{R} &= & \frac{2}{n}
\sum_{k=1}^{n-1}
\alpha_{k/n} \sin(\pi k/n) \, C_{k/n}\,
=
 2\sum_{k=1}^{n-1}
\alpha_{k/n}  \sin(\pi k/n) \, E_{k/n}\,.
\end{eqnarray}
Writing the quadratic form in the exponent of the generic term of the sum defining $\Theta(0 | T )$, 
it is straightforward to see that (\ref{Fn2}) is invariant under $\eta \leftrightarrow 1/\eta$.
The matrices (\ref{I matrix def}) and (\ref{R matrix def}) are respectively the imaginary and the real part of the  matrix
\begin{equation}\fl 
\label{period matrix}
\tau =
\mathcal{R} + i \mathcal{I}  =
\frac{2}{n} \sum_{k=1}^{n-1} \tau_{k/n} \sin(\pi k/n) \, C_{k/n} =
2 \sum_{k =1}^{n-1} \tau_{k/n} \sin(\pi k/n) \, E_{k/n}\,.
\end{equation}
We can write $\Theta(0 | T)$ in (\ref{Fn2}) also as
\begin{equation}
\label{Theta with tau}
\Theta(0 | T )=\sum_{{\bf p}, \tilde{\bf p}}
\exp 
\big[\,
i \pi \big( {\bf p}^{\rm t} \cdot \tau \cdot {\bf p} -  { \tilde{\bf p}}^{\rm t} \cdot \bar{\tau} \cdot { \tilde{\bf p}} \big)\,
\big]
\end{equation}
where
\begin{equation}
({\bf p}, \tilde{\bf p}) =
\left( \frac{{\bf n}}{\sqrt{2\eta}}+\frac{{\bf m} \sqrt{2\eta}}{2},
\frac{{\bf n}}{\sqrt{2\eta}}-\frac{{\bf m} \sqrt{2\eta}}{2} \right),
\qquad
{\bf n} , {\bf m} \in  \mathbb{Z}^{n-1}.
\end{equation}
Comparing with the results in the literature \cite{Dijkgraaf:1987vp}, from (\ref{Theta with tau}) we conclude that $\tau$ is 
the period matrix of the genus $g=n-1$ Riemann surface we are dealing with.
For $x \in (0,1)$ we have that $\alpha_{k/n} =0$ and therefore all the elements of the matrix $\mathcal{R}$ vanish identically. 
In this case $\Theta(0 | T) = \Theta(0 |  i \eta \mathcal{I} ) \Theta(0 | i\mathcal{I} /\eta) $ and the  $\eta \leftrightarrow 1/\eta$ 
invariant expression found in \cite{cct-09} is recovered.

\subsubsection{The special case $n=2$ for the compactified boson.}

It is instructive to analyze the details of the case $n=2$ with complex $x$. 
Now the Riemann surface is a torus and the period matrix reduces to a number: its modulus $\tau_{1/2}(x)$. First we recall that
\begin{equation}
\label{Fkn n=2}
F_{1/2}(x) = \frac{2}{\pi}\,K(x) = \theta_3(\tau_{1/2})^2\,,
\end{equation}
where $K(x)$ is the complete elliptic integral of the first kind and
\begin{equation}
\label{xtau n=2}
\tau_{1/2}(x) = {i}\,\frac{K(1-x)}{K(x)}\,,
\qquad
x(\tau_{1/2}) = \left(\frac{\theta_2(\tau_{1/2})}{\theta_3(\tau_{1/2})}\right)^4\,.
\end{equation}
For $n=2$ we have that $ \mathcal{T}_2 = \overline{ \mathcal{T}}_2$. From Eqs. (\ref{4ptgen}), (\ref{Fn2}), 
and (\ref{Fkn n=2}) we find that the four point function of the twist fields ${\cal T}_2$ for the compactified real boson reads 
\begin{eqnarray}\fl
\label{4point n=2 one}
\langle \mathcal{T}_2(z_1) \mathcal{T}_2(z_2) \mathcal{T}_2(z_3) \mathcal{T}_2(z_4)  \rangle 
& = &
\left| \frac{z_{31}  z_{42}}{z_{21} z_{32} z_{41}  z_{43}}\right|^{1/4}
\frac{\Theta(0|T)}{\big|\theta_3(\tau_{1/2})^2\big|} \nonumber
\\ \fl 
\label{4point n=2 one v1}
&=&
\left| \frac{1}{z_{21} z_{31}  z_{41}  z_{32}  z_{42}  z_{43} [x(1-x)]^2}\right|^{1/12}
\frac{\Theta(0|T)}{\big|\theta_3(\tau_{1/2})^2\big|}\,,
\end{eqnarray}
where $T$ is the $2\times 2$ matrix  given by (\ref{matrix T}) for $n=2$, namely
\begin{equation}
T =\left(\begin{array}{cc}
i \eta \beta_{1/2} & \alpha_{1/2} \\
 \alpha_{1/2}  & {i} \,\beta_{1/2}/\eta
\end{array}\right)\,.
\end{equation}
From (\ref{xtau n=2}) and the identity $\theta_4^4=\theta_3^4-\theta_2^4$ one can find that
\begin{equation}
\label{x(1-x) identity}
x(1-x) =\left(\frac{\theta_2(\tau_{1/2})\,\theta_4(\tau_{1/2})}{\theta_3(\tau_{1/2})^2}\right)^4\,.
\end{equation}
The transformation properties of the Jacobi theta functions lead also to the following identities 
(here $\tau=\tau_{1/2}$)
\begin{equation}\fl 
\label{SL2Z trans}
x(-{1}/{\tau}) = 1-x(\tau),
\qquad
x\Big(\frac{-\tau}{\tau-1}\Big) = \frac{1}{x(\tau)},
\qquad
x(\tau+1) = \frac{x(\tau)}{x(\tau)-1}.
\end{equation}
Plugging (\ref{x(1-x) identity}) into (\ref{4point n=2 one v1}), we find 
\begin{eqnarray}\fl
\langle \mathcal{T}_2(z_1) \mathcal{T}_2(z_2) \mathcal{T}_2(z_3) \mathcal{T}_2(z_4)  \rangle 
& = &
\left| \frac{1}{z_{21}  z_{31}  z_{41}  z_{32}  z_{42}  z_{43} }\right|^{1/12}
\frac{\Theta(0|T)}{|\theta_2(\tau_{1/2}) \theta_3(\tau_{1/2})\theta_4(\tau_{1/2})|^{2/3}} \nonumber
\\ \fl 
\label{4point n=2 last}
&=&
\left| \frac{1}{z_{21}  z_{31}  z_{41}  z_{32}  z_{42}  z_{43}  }\right|^{1/12}
\frac{\Theta(0|T)}{2^{1/3} \big|\eta(\tau_{1/2})\big|^2}\,,
\end{eqnarray}
where in the last step we used the identity $2 \eta^3 = \theta_2 \theta_3 \theta_4$
and $\eta(z)$ is the Dedekind eta function, which should not
be confused with the compactification parameter.
Now we observe that $\mathcal{Z}_2(\tau_{1/2})\equiv \Theta(0|T)/|\eta(\tau_{1/2}) |^2$ is the partition function of the 
compactified boson on the torus (see e.g. Eq. (10.62) of \cite{cft-book} with $\eta_{\rm here}=R^2_{\rm there}/2$).

Let us consider the exchanges $z_1 \leftrightarrow z_3$, $z_2 \leftrightarrow z_3$ and $z_3 \leftrightarrow z_4$ separately. 
From (\ref{xgen}) it is easy to see that they correspond to the following involutions: $x \leftrightarrow 1-x$, $x \leftrightarrow 1/x$ and 
$x \leftrightarrow x/(x-1)$ respectively. 
From (\ref{SL2Z trans}), one notices that they are also associated some $SL(2,\mathbb{Z})$ transformations of $\tau_{1/2}$.
Since the prefactor containing the $z_{ij}$'s in (\ref{4point n=2 last}) is invariant under the three exchanges above and the partition function $\mathcal{Z}_2(\tau_{1/2})$ is $SL(2,\mathbb{Z})$ invariant \cite{cft-book}, we conclude that the four point function (\ref{4point n=2 last}) is invariant under $z_1 \leftrightarrow z_3$, $z_2 \leftrightarrow z_3$ and $z_3 \leftrightarrow z_4$ separately.
This allows us to find that the ratio defined in (\ref{Rndef}) for the compactified boson is given by
\begin{equation}
R_2(y) = 1
\end{equation}
identically for $0<y<1$. However, we stress that, ultimately, this is just the fact that $\mathcal{T}_{2} = \bar{\mathcal{T}}_{2}$.

\section{Systems with boundaries}
\label{bou0}

Let us consider a 1D  system on the semi-infinite line, say $[0,\infty)$ and a bipartition 
$A_2=[0,\ell]$ and $A_1$ the reminder. 
We then have 
\bea
\Tr (\rho_A^{T_2})^{n_e}= \langle {\cal T}_{n_e}^2(\ell) \rangle=
\tilde{c}_{n_e/2}^2 \Big(\frac{2\ell}a\Big)^{-{c}/{6}({n_e}/2-2/{n_e})}\,,
\nonumber \\ 
\Tr (\rho_A^{T_2})^{n_o}=  \langle {\cal T}_{n_o}^2(\ell) \rangle=
\tilde{c}_{n_o} \Big(\frac{2\ell}a\Big)^{-{c}/{12}(n_o-1/{n_o})},
\label{1intobccft}\\
{\cal E}=\frac{c}4 \ln \frac{2\ell}a+2\ln \tilde{c}_{1/2}.\nonumber
\eea
The non-universal constants $\tilde{c}_n$ are not the same as those appearing in the case of the 
infinite system $c_n$, but are related to them via the Affleck-Ludwig boundary entropy  \cite{al-91}
as discussed in Refs. \cite{cc-04,zbfs-06}.

Like for a periodic system, in the case of a finite system of length $L$ with the same boundary conditions on both ends, 
the negativity can be obtained by a simple logarithmic mapping which results in the net 
replacement $\ell\to \frac{L}\pi\sin (\pi\ell/L)$ and we get 
\be
{\cal E}=\frac{c}4 \ln \left[ \frac{2L}{\pi a}\sin \Big(\frac{\pi\ell}L \Big)\right]+2\ln \tilde{c}_{1/2}.
\label{1intobcfin}
\ee

The case of two adjacent finite intervals (or even one interval not adjacent to a boundary) 
is a two-point function in the half-plane, which has the same complexity as 
a  four-point function in the full plane and then depends on the full operator content and not only on the central 
charge, as  for the case of two disjoint intervals on the full line.
However, some results can be achieved in a quite general way.

Let us first consider the case of two adjacent intervals, the first one starting from the boundary, 
i.e. $A_1=[0,\ell_1]$ and $A_2=[\ell_1+\ell_2]$ where $B=[\ell_1+\ell_2,\infty)$ is the remainder.  
We place the spatial coordinate along the imaginary direction in the complex plane,
while the imaginary time is on the real direction, i.e. $z=\tau+i x$, thus the system lies on the 
upper half plane (UHP). 
The traces of integer powers of the partial transpose reduced density matrix,  
correspond to the UHP two-point function
\be\fl
\Tr (\rho_A^{T_2})^n=\langle \overline {\cal T}_n^2(z_1)  {\cal T}_n(z_2)\rangle_{\rm UHP}\,,
\qquad 
{\rm with}\; z_1=i\ell_1,\;\; z_2=i(\ell_1+\ell_2)\,. 
\ee
The general form of such two-point function can be obtained by images method and it is \cite{c-84}
\be\fl 
\langle \overline {\cal T}_n^2(z_1)  {\cal T}_n(z_2)\rangle_{\rm UHP}\propto
\frac{{\cal F}^b_n(y)}{|z_{1\bar 1}|^{\Delta_{{\cal T}_n^2}} |z_{2\bar 2}|^{\Delta_{{\cal T}_n}}}
\,, \qquad y=\frac{z_{12}z_{\bar 1\bar 2}}{z_{1\bar2}z_{\bar1 2}}\,,
\ee
where $z_{\bar i}=\bar{z}_i$, $z_{ij}=z_i-z_j$, 
$y$ is the four point ratio built  with the two points and their images, and 
${\cal F}_n^b(y)$ is a scale invariant function of $y$. 
Calculating this function is   complicated, 
but there are some general conclusions which can be drawn  without 
making any calculation. 
Indeed, let us assume $\ell_1=\ell_2=\ell$. 
In this case $y=1/9$ is constant, independent of $\ell$, as the physical intuition suggests.
Then we have 
\be
\langle \overline {\cal T}_n^2(i\ell)  {\cal T}_n(i2\ell)\rangle_{\rm UHP}\propto
\frac1{\ell^{\Delta_{{\cal T}_n^2}} \ell^{\Delta_{{\cal T}_n}}}\,. 
\ee
Using the values of $\Delta_{{\cal T}_n^2}$ in Eqs. (\ref{Dtne}) and (\ref{Dtno}), we have
\be
\Tr (\rho_A^{T_2})^n\propto \left\{
\begin{array}{ll}
\ell^{-c/6(n-1/n)} & n\; {\rm odd}\,,\\
\\
\ell^{-c/12(n-1/n) -c/6(n/2-2/n)} & n\; {\rm even}\,.
\end{array}
\right.
\label{2adjbound}
\ee
In particular, the $n$ dependence for $n=n_e$ even is quite peculiar, being 
 different from others found before. 
In the limit $n_e\to1$, we obtain
\be
{\cal E}=\frac{c}4 \ln \ell+{\rm const}\,,
\ee
which however is  similar to what  found in other circumstances.
We stress that in the case of a finite system of length $L$, the 
result above applies if and only if one scales $L$ with $\ell$ in such a way to keep 
fixed the four-point ratio $y$. In all other circumstances, one should use the 
full formula with the unknown function ${\cal F}_n^b(y)$.

As a last example we  mention the case of $A_2$ being a single interval 
detached from the boundary and $A_1$ the remaining part of the half-line. 
In such a case, $\rho_A$ corresponds to a pure system and so the general result
${\cal E}=S^{(1/2)}_{A_2}$ in Eq. (\ref{genEpure}) applies and
 the results for the R\'enyi entropies discussed in Refs. \cite{cct-09,fc-11}
apply straightforwardly.

\section{The harmonic chain}

In this section we present accurate numerical checks of our CFT predictions for the harmonic chain 
with periodic and Dirichlet  boundary conditions. 
The Hamiltonian of the  harmonic chain with $N$ lattice sites and with nearest neighbor interaction 
(but the procedure is easily generalized to arbitrary interactions)  is
\begin{equation}
\label{hamiltonian chain periodic}
H= \sum_{n=0}^{N-1}\left[\,
\frac{1}{2M}p_n^2+\frac{M \omega^2}{2}q_n^2+
\frac{K}{2} (q_{n+1}-q_n)^2\right]\,,
\end{equation}
where periodic boundary conditions correspond to $q_{N}= q_0$ (and $p_{N}= p_0$) while
Dirichlet  boundary conditions to $q_0=q_{N}=p_0=p_{N}=0$  (these are sometimes called 
fixed wall conditions).
The variables $p_n$ and $q_n$ satisfy standard commutation relations $[q_n, q_m]=[p_n,p_m]=0$
and $[q_n,p_m]= i\delta_{nm}$. 

The chain is defined by the three parameters $\omega,K,M$, however not all of them 
are essential. Indeed, we can use a canonical transformation 
\be 
(q_n, p_n) \rightarrow ((MK)^{1/4} q_n, p_n/(MK)^{1/4}) ,
\ee
 to eliminate one of the three parameters. 
Introducing $a=\sqrt{M/K}$ then we can rewrite the Hamiltonian as
\begin{equation}
\label{hcp2}
H= \sum_{n=0}^{N-1}\left[\,
\frac{1}{2a}p_n^2+\frac{a \omega^2}{2}q_n^2+
\frac{1}{2a} (q_{n+1}-q_n)^2\right]\,.
\end{equation}
In this form, it is evident that the Hamiltonian is the  lattice discretization of a 
free-boson (Klein Gordon field) with lattice spacing $a$ and mass $\omega$.  
Indeed in the limit $a\to0$, $N\to\infty$, with $Na=L$, we can replace
\be
q_n\to \phi(x), \qquad \frac{p_n}a \to \pi(x)=\dot{\phi}(x), \qquad {\rm with }\;\, x=n a,
\ee
(satisfying $[\phi(x),\pi(x')]= i\delta(x-x')$ with $\delta_{n,n'}\to a \delta(x-x')$) and the Hamiltonian above 
reduces to the two-dimensional euclidean action
\be
S= \frac12\int_0^L dx\int d\tau\left[(\partial_\mu \phi)^2+\omega^2\phi^2\right]\,.
\ee
For $\omega=0$ the theory is critical and conformal with central charge the $c=1$.
The boson field $\phi(x)$ is not compactified and so the theory corresponds
to the limit $\eta\to\infty$ in the previous CFT description.  

\subsection{Diagonalization and correlation functions of the harmonic chain with periodic boundary conditions.}

For periodic boundary conditions, the Hamiltonian (\ref{hamiltonian chain periodic}) can be diagonalized 
by exploiting the translational invariance and introducing the Fourier transforms of the canonical variables, namely
\be
q_r =\frac{1}{\sqrt{N}} \sum_{k=0}^{N-1} \tilde{q}_k \,e^{2\pi i kr/N},
\qquad
\tilde{q}_k =\frac{1}{\sqrt{N}} \sum_{s=0}^{N-1} q_s \,e^{-2\pi i ks/N},
\ee
(and similarly for $p_r$) for $r=1, \dots, N$.
The Hamiltonian (\ref{hamiltonian chain periodic}) then becomes
\begin{equation}
\label{H fourier periodic}
H =
\sum_{k=0}^{N-1} \left(
\frac{1}{2M}\,\tilde{p}_k \tilde{p}_{N-k}+
\frac{M \omega_k^2}{2} \tilde{q}_k \tilde{q}_{N-k}
\right),
\end{equation}
where
\begin{equation}
\label{omegak}
\omega_k \equiv 
\sqrt{\omega^2+\frac{4K}{M} \sin\Big(\frac{\pi k}N\Big)^2} \,\geqslant\, \omega,
\hspace{1.5cm} 
k=0,\dots, N-1.
\end{equation}
Notice that $\omega_{-k} = \omega_{N-k} =\omega_{k}$. As usual,  we identify 
$\tilde{p}_{N-k}$ with $\tilde{p}_{-k}$ and $\tilde{q}_{N-k}$ with $\tilde{q}_{-k}$.
The minimum values assumed by the dispersion relation $\omega_k$'s   is $\omega_0 = \omega$.
From the canonical commutation relation $[ q_r  , p_s] = i \delta_{rs}$, one finds that $\tilde{q}_k$ 
and $\tilde{p}_{N-k}$ are canonically conjugate (i.e. $[\tilde{q}_k  , \tilde{p}_{-k'}] = i \delta_{k, k'}$).
Now one defines the annihilation and creation operators as 
\begin{equation}\fl
\label{a adag periodic}
a_k \equiv \sqrt{\frac{M \omega_k}{2}} \left( \tilde{q}_k+\frac{i}{M \omega_k}\, \tilde{p}_k\right),
\qquad
a_k^\dagger \equiv \sqrt{\frac{M \omega_k}{2}} \left( \tilde{q}_{-k}-\frac{i}{M \omega_k}\, \tilde{p}_{-k}\right),
\end{equation}
which satisfy the algebra $[a_k  , a_{k'}] = [a_k^\dagger  , a_{k'}^\dagger] = 0$ and $ [a_k  , a^\dagger_{k'}]= i \delta_{k, k'}$.
Then the Hamiltonian (\ref{H fourier periodic}) becomes
\begin{equation}
\label{H a adagger}
H =
\sum_{k=0}^{N-1}  \omega_k
\left(
a_k^\dagger a_k +\frac{1}{2}
\right).
\end{equation}
The ground state of the harmonic chain is then the vacuum $|0\rangle$ of the $a_k$, satisfying $a_k|0\rangle=0$ for 
any $k$.

The two point functions $\langle q_r q_s \rangle $ and $\langle p_r p_s \rangle $
in the vacuum are worked out writing $(\tilde{q}_k ,\tilde{p}_k)$ in terms of the operators $a_k$ and $a_k^\dagger$, finding
\begin{eqnarray}
\label{qq corr}
\langle 0| q_r q_s  |0\rangle &=&
\frac{1}{2N}  \sum_{k=0}^{N-1} \frac{1}{M \omega_k} \cos\Big[\frac{2\pi k(r-s)}N\Big],
\\
\langle 0| p_r p_s  |0\rangle &=&
\frac{1}{2N}  \sum_{k=0}^{N-1} M \omega_k \cos\Big[\frac{2\pi k(r-s)}N\Big]. \nonumber
\end{eqnarray}
These correlation functions can be organized in correlation matrices  $\mathbb Q$ and $\mathbb P$ whose elements are 
the correlation functions, i.e.
\be
\mathbb Q_{rs} \equiv \langle 0| q_r q_s  |0\rangle,  \qquad \mathbb P_{rs} \equiv \langle 0| p_r p_s  |0\rangle, 
\label{PQdef}
\ee
satisfying $\mathbb Q\cdot \mathbb P = {\mathbb I}_N/4$, where ${\mathbb I}_N$ is the $N \times N$ identity matrix.
We also have $\langle 0| q_r p_s  |0\rangle =  i \delta_{r, s }/2$.

It is very important for what follows to stress that 
$\langle 0| q_r q_s  |0\rangle$ is not well defined when $\omega=0$ because  $\omega_0=0$.
This is the well known problem arising from the {\it zero mode}, i.e. a constant translational invariant field 
configuration.  
For $\omega >0$ we can isolate the term in Eq. (\ref{qq corr}) which diverges as $\omega \to 0$ 
\begin{equation}
\label{qq corr sep}
\langle 0| q_r q_s  |0\rangle =
 \frac{1}{2N M \omega} +
\frac{1}{2N}  \sum_{k=1}^{N-1} \frac{1}{M \omega_k}\cos\Big[\frac{2\pi k(r-s)}N\Big],
\end{equation}
and the remaining sum is finite also in the limit $\omega\to0$.
For $\omega=0$, the zero mode leads to divergent expressions, thus 
we work at finite but small $\omega$ such that $\omega L\ll 1$.

\subsection{Diagonalization and correlation functions of the harmonic chain with Dirichlet boundary condition.}

For the Dirichlet   boundary conditions, the diagonalization cannot be performed by Fourier transform because of 
the breaking of translational invariance. However, we can simply use the Fourier sine transform, 
defining as in Ref. \cite{lsv-08}
\begin{equation}\fl
\tilde{q}_k =\sqrt{\frac{2}{N}}  \sum_{r=1}^{N-1} q_r   \sin\bigg(\frac{\pi k r}{N}\bigg),
\qquad
q_r =\sqrt{\frac{2}{N}}  \sum_{k=1}^{N-1} \tilde{q}_k   \sin\bigg(\frac{\pi k r}{N}\bigg),
\end{equation}
(and similarly for $p_r$) with $r=1, \dots, N-1$. In this basis the Hamiltonian is diagonal
\begin{equation}
\label{H fourier fwbc}
H =
\sum_{k=1}^{N-1} \left(
\frac{1}{2M}\,\tilde{p}^2_k+
\frac{M \tilde{\omega}_k^2}{2} \tilde{q}^2_k 
\right),
\end{equation}
where the dispersion relation reads
\begin{equation}
\label{omegak fwbc}
\tilde{\omega}_k \,\equiv\,
\sqrt{ \omega^2+\frac{4K}{M}  \sin^2\bigg(\frac{\pi k}{2 N}\bigg)} > \omega,
\qquad
k= 1,\dots, N-1.
\end{equation}
The key difference with respect to the periodic case is that $\tilde{\omega}_k  > \omega \geqslant 0$ for $k=1,\dots, N-1$.
Thus, in this model all the quantities are well defined  even for $\omega =0$.
In terms of the  creation and the annihilation operators,  
the Hamiltonian takes the form  (\ref{H a adagger}) but with $\tilde{\omega}_k$ replacing $\omega_k$.

The correlators  in the ground-state of the Hamiltonian are
\begin{eqnarray}
\langle 0| q_r q_s  |0\rangle &=&
\frac{1}{N}  \sum_{k=1}^{N-1} \frac{1}{M \tilde{\omega}_k} 
\sin\Big(\frac{\pi k r}{N}\Big) \sin\Big(\frac{\pi k s}{N}\Big) =\mathbb Q_{rs}, \nonumber
\\
\label{pp corr fwbc}
\langle 0| p_r p_s  |0\rangle &=&
\frac1{N}  \sum_{k=1}^{N-1} M \tilde{\omega}_k  
\sin\Big(\frac{\pi k r}{N}\Big) \sin\Big(\frac{\pi k s}{N}\Big) =\mathbb P_{rs},
\end{eqnarray}
where $r,s =1, \dots , N$. 

For these Dirichlet boundary conditions we can write down exact closed formulas for 
$\omega =0$ and in the thermodynamic limit $N \rightarrow \infty$, when  
the correlators (\ref{pp corr fwbc}) become 
\begin{eqnarray}\fl
\nonumber
\langle 0| q_r q_s  |0\rangle
& = &
\frac{1}{2\pi \sqrt{MK}} 
\Big( \psi(1/2+r+s) - \psi(1/2+r-s) \Big),
\\ \fl 
\label{pp corr fwbc N=inf om=0}
\langle 0| p_r p_s  |0\rangle
&=&
\frac{2\sqrt{MK}}{\pi} 
\Big(
\frac{1}
{4(r+s)^2-1}
-
\frac{1}
{4(r-s)^2-1}
\Big),
\end{eqnarray}
where $\psi(z)$ is the digamma function.

\subsection{The reduced density matrix and its partial transpose.}

The construction of the reduced density matrix $\rho_A$  for the ground-state of the
harmonic chain has been detailed in a several papers
with slightly different theoretical approaches \cite{pc-99,Audenaert02,p-03,br-04,area-06,pe-09}.
Following Refs. \cite{p-03,pe-09}, we can easily relate $\rho_A$ to the correlation matrices $\mathbb Q_{rs}=\langle q_rq_s\rangle$ 
and $\mathbb P_{rs}=\langle p_r p_s\rangle$ (derived above for periodic and Dirichlet boundary conditions) restricted to 
the part $A$ of the system both if the subsystem is connected or formed by an arbitrary number of  pieces.
Indeed, since the ground-state is Gaussian in normal coordinates (provided that no normal frequency vanishes)
the reduced density matrix for an arbitrary subsystem $A$ can be written in the Gaussian form \cite{p-03,pe-09}
\be
\rho_A\propto \exp\Big (-\sum_{j\in A} \epsilon_j b^\dagger_j b_j\Big),
\label{gaus}
\ee  
where $b^\dagger_j$ and $b_j$ are bosonic creation and annihilation operators
related to the original operator in the subsystem by a canonical transformation. 
At this point it is straightforward to fix the value of the $\epsilon_j$ by imposing that the 
correlation functions of $q$'s and $p$'s in $A$ calculated with $\rho_A$ equal the ground state ones (e.g. the one 
calculated above for periodic and Dirichlet boundary conditions).
Denoting with $\mathbb P_A$ and $\mathbb Q_A$ the restriction of the matrices 
$\mathbb P$ and $\mathbb Q$ to the subsystem $A$ and with 
$\mu^2_j$ ($j=1\dots  \ell$, with $\ell$ the number of sites in $A$, not necessarily connected) 
 the eigenvalues of the matrix ${\mathbb P_A\cdot \mathbb Q_A}$, i.e.
\be
{\rm Spectrum}(\mathbb Q_A\cdot \mathbb P_A)=\{\mu_1^2,\dots \mu_\ell^2\}\,,
\ee
one finds $\mu_j=\frac12 \coth\frac{\epsilon_j}2$ \cite{p-03}. 
(Alternatively one could work with the $2\ell\times 2\ell$ matrix
formed by two-point functions of position and momentum operators, 
called reduced covariance matrix. Its reduction to a diagonal form is related to the problem of finding the 
symplectic spectrum of a symmetric and positive definite matrix \cite{Audenaert02,br-04}.)

From the eigenvalues $\mu_j$, using the explicit form of $\rho_A$ above, we obtain the R\'enyi entropies
\begin{equation}
\label{renyi harmonic chain}
\Tr \rho_A^n =\prod_{j=1}^\ell
\Bigg[
\bigg(\mu_j+\frac{1}{2}\bigg)^n - \bigg(\mu_j-\frac{1}{2}\bigg)^n \Bigg]^{-1} ,
\end{equation} 
and the entanglement entropy as
\begin{equation}
\label{EE harmonic chain}
S_A
=\sum_{j=1}^\ell
\bigg[
\bigg(\mu_j+\frac{1}{2}\bigg) \ln \bigg(\mu_j+\frac{1}{2}\bigg) 
- 
\bigg(\mu_j-\frac{1}{2}\bigg) \ln \bigg(\mu_j-\frac{1}{2}\bigg) 
\bigg].
\end{equation} 

The partial transpose of the reduced density matrix has been constructed in Ref.~\cite{Audenaert02},
where it has been shown that the partial transposition with respect to the subsubsytem $A_2$ 
has the net effect of changing the sign of the momenta corresponding to the subsystem $A_2$, 
leaving the partial transpose reduced density matrix $\rho_A^{T_2}$ a Gaussian matrix of the 
form (\ref{gaus}).
This means that we can simply replace the matrix $\mathbb P_A$ with a matrix in which all the 
momenta in $A_2$ have been changed sign, i.e. 
\be
\mathbb P_A^{T_2}= \mathbb R_{A_2} \mathbb P_A \mathbb R_{A_2}\,,
\label{Pt2}
\ee
where $\mathbb R_{A_2}$ is the $\ell \times \ell$ diagonal matrix having $-1$ in correspondence of the sites belonging to 
$A_2$ and $+1$ otherwise, i.e. $(\mathbb R_{A_2} )_{rs} = \delta_{rs}(-1)^{\delta_{r \in A_2}}$. 
Notice that $\mathbb P^{T_1}_A = \mathbb P^{T_2}_A$ because $\mathbb R_{A_1} = -\mathbb R_{A_2}$.
Then the eigenvalues $\nu_j$ of the product $\mathbb Q_A \cdot \mathbb P_A^{T_2}$ determine the 
full spectrum of $\rho_A^{T_2}$.  Denoting 
\begin{equation}
\textrm{Spectrum} \big( \mathbb Q_A \cdot \mathbb P_A^{T_2}  \big)=\{ \nu_1^2, \dots , \nu_\ell^2 \},
\end{equation} 
we have 
\begin{equation}
\label{traces partial transpose chain}
\textrm{Tr} \big(\rho_A^{T_2}\big)^n
=\prod_{j=1}^{\ell}
\Bigg[
\bigg(\nu_j+\frac{1}{2}\bigg)^n - \bigg(\nu_j-\frac{1}{2}\bigg)^n\Bigg]^{-1} ,
\end{equation} 
and the  trace norm 
\begin{equation}
\label{norm partial transpose chain}
|| \rho_A^{T_2} ||
=\prod_{j=1}^{\ell}
\Bigg[
\bigg|\nu_j+\frac{1}{2}\bigg| -  \bigg|\nu_j-\frac{1}{2}\bigg| \Bigg]^{-1} 
=\prod_{j=1}^{\ell}
\textrm{max} \bigg[1,\frac{1}{2\nu_j} \bigg].
\end{equation} 
From this quantity we easily get the  negativity $\mathcal{N}$ and the logarithmic negativity ${\cal E}$.
Notice that only $\nu_j < 1/2$ contribute to (\ref{norm partial transpose chain}).

Let us then summarize the practical procedure to find the negativity and $\Tr(\rho_A^{T_2})^n$ for an 
arbitrary tripartion $A_1\cup A_2 \cup B$ in the ground-state of an harmonic chain:
\begin{itemize}
\item From the correlation matrices $\mathbb Q$ and $\mathbb P$ (which for periodic and Dirichlet boundary 
conditions are given in Eqs. (\ref{qq corr}) and (\ref{pp corr fwbc}) respectively), construct the reduced correlation matrices 
$\mathbb Q_A$ and $\mathbb P_A$ by erasing the rows and columns corresponding to the part $B$ of the system.
\item Change the signs of the momenta in the part $A_2$ to construct the matrix ${\mathbb P}_A^{T_2}$ as in Eq. (\ref{Pt2}).
\item Calculate the eigenvalues $\nu_j^2$ of the product $\mathbb Q_A\cdot {\mathbb P}_A^{T_2}$.
\item Use Eq. (\ref{traces partial transpose chain}) to  calculate $\Tr(\rho_A^{T_2})^n$ and Eq. (\ref{norm partial transpose chain}) 
to obtain the negativity from $\nu_j$. 
\end{itemize}

\begin{figure}[t]
\includegraphics[width=\textwidth]{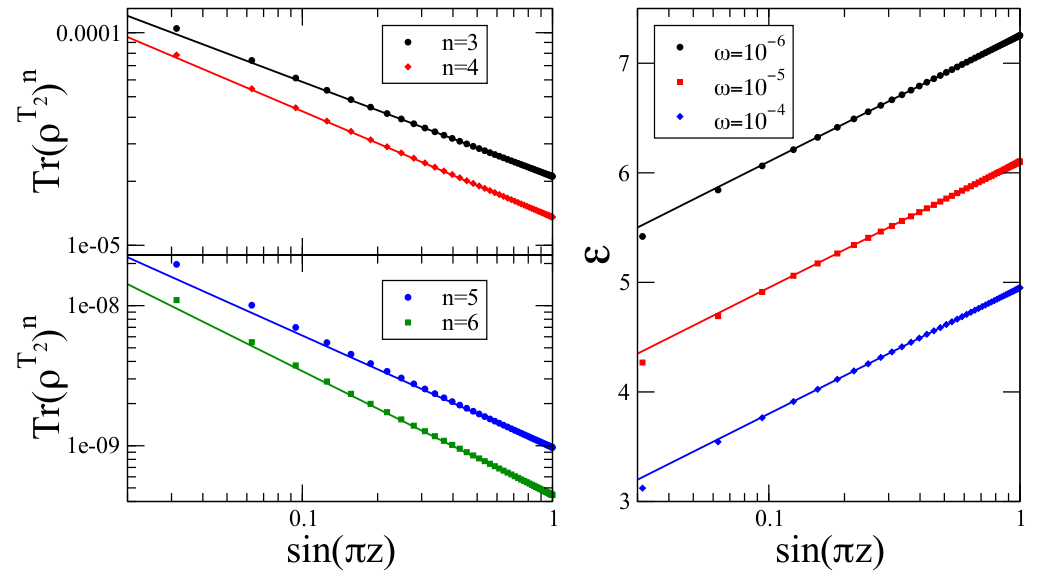}
\caption{Entanglement in a periodic chain of total length $L=100$. We consider the bipartition 
between an interval of length $\ell$ and the remainder. Here $z=\ell/L$.
Left: $\Tr (\rho^{T_2})^n$ for $n=3,4,5,6$ as function of $\sin(\pi z)$. 
The straight lines are the CFT predictions where the only free parameter is the overall amplitude 
$c_n$ (cf. Eqs. (\ref{cftTr1int2})  and  (\ref{cftTr1int})).
Right: The logarithmic negativity ${\cal E}$ vs the CFT prediction (\ref{EN1int}).
The three curves correspond to different values of $\omega$ and show the influence of the zero mode.
Indeed they are parallel (in log-linear scale) and the zero mode affects only the non-universal additive 
constant.
}
\label{1intpbc}
\end{figure}

In the semi-analytic calculations presented in the following we will always work in 
units $M=K=a=1$ and so the total length is $L=N a=N$.

\subsection{The negativity for one interval.}

We consider here the case when $A_1$ is a block of $\ell$ consecutive sites and $A_2$ the remainder, i.e.  
 $B\to\emptyset$, as shown in the bottom of Fig. \ref{intervals}. 
In this case, the negativity and the traces $\Tr (\rho^{T_2})^n$ 
coincide with the R\'enyi entropies for any pure state (cf. Eq. (\ref{pureqm})) as proved in Sec. (\ref{secpure}).
It is however important to report these results for a twofold reason.
On the one hand, knowing a priori the final result of the calculation provides a
non-trivial check of the numerical procedure which indeed differs substantially 
when calculating $\Tr (\rho^{T_2})^n$ and $\Tr \rho_{2}^n$ since for latter 
there is no need neither of a partial transposition, nor of the correlation matrices in the 
part $A_1$ 
(that involves basically the same procedure for one and two intervals).
On the other hand, these controlled calculations give also an idea of the corrections 
to the scaling to the asymptotic CFT formulas.

In the case of periodic boundary conditions, we cannot work directly 
with $\omega=0$, because of  the zero mode. 
Thus, we consider several values of $\omega$ imposing the 
further constraint $\omega L\ll 1$ to ensure that all the data for any $1<\ell<L$ are in the conformal regime.  
The data for $\Tr(\rho^{T_2})^n$ in the periodic harmonic chain are reported in Fig. \ref{1intpbc} 
for $n=3,4,5,6$. 
The agreement with the conformal predictions in Eq. (\ref{cftTr1int}) is excellent. 
The visible deviations for small $\ell$ are the corrections
to the scaling to the entanglement R\'enyi entropy discussed in Ref. \cite{un1,un2,ce-10} of
the unusual form $\ell^{-2/n_R}$ where $n_R$ is the index of the corresponding 
R\'enyi entropy (i.e. $n_R=n$ for $n$ odd and $n_R=n/2$ for $n$ even).

\begin{figure}[t]
\includegraphics[width=\textwidth]{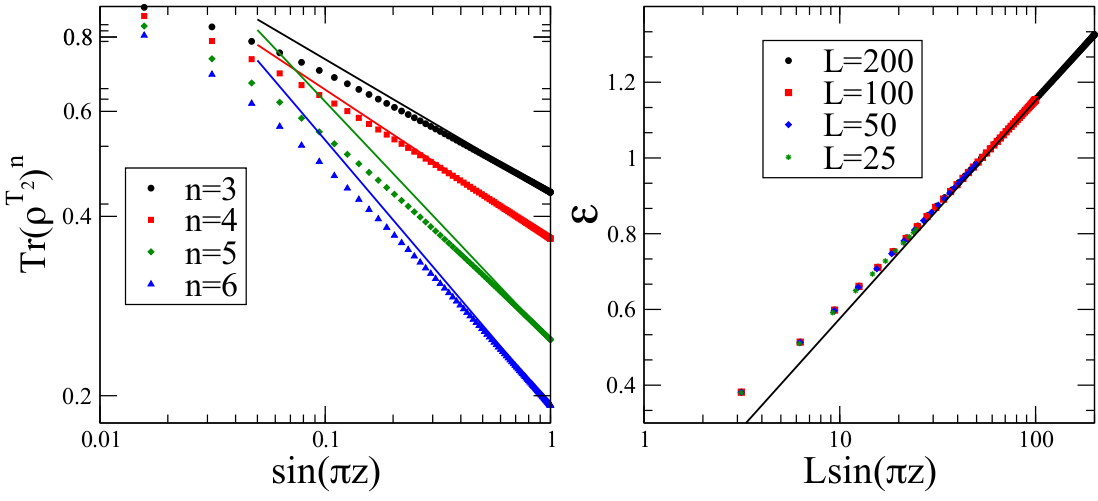}
\caption{Entanglement in a critical ($\omega=0$) harmonic  chain with Dirichlet boundary conditions of total length $L$. 
We consider the bipartition between the interval  $A_1=[0,\ell]$ and the remainder $A_2=[\ell,L]$. 
Here $z=\ell/L$.
Left: $\Tr (\rho^{T_2})^n$ for $n=3,4,5,6$ and for $L=200$ as function of $\sin(\pi z)$. 
The straight lines are the CFT predictions where the only free parameter is the overall amplitude 
$\tilde{c}_n$ (cf. Eq. (\ref{1intobccft})).
Right: The logarithmic negativity ${\cal E}$ for $L=25,50,100,200$.
The four sets of data collapse on the same curve when plotted against the chord length, 
as they should.
}
\label{1intdbc}
\end{figure}

In Fig. \ref{1intpbc} we report also the logarithmic negativity as function of the chord length,
finding a perfect agreement with the conformal prediction (\ref{EN1int}). 
We report the data for three different values of $\omega$, all satisfying $\omega L\ll1$ and
so in the conformal regime. It is evident that in logarithmic-linear scale the three curves are parallel
confirming that the zero mode only affects the value of the non-universal additive constant and not the 
leading logarithmic behavior of ${\cal E}$ with the subsystem sizes. 

We then perform the same analysis for a finite system of total length $L$ with 
Dirichlet boundary conditions considering the interval $A_1$ starting from the boundary 
up to $\ell$ and $A_2$ the remaining $L-\ell$ sites. 
In this case, as discussed above, there is no zero mode and we can work directly at $\omega=0$. 
The data for $\Tr(\rho^{T_2})^n$ are reported in Fig. \ref{1intdbc} for several values of $n$, all for  $L=200$.
It is evident that increasing $\ell$ the data approach the CFT predictions in Eq. (\ref{1intobccft}), but in this case 
the corrections to the scaling are much larger than in the periodic case. 
This does not come unexpected, indeed in Refs. \cite{un1,un2,fc-11} 
it has been shown that in the presence of the boundaries the corrections 
to the asymptotic results  are of the form $\ell^{-1/n_R}$ (where again $n_R$ is the index of the corresponding 
R\'enyi entropy), i.e. they have an exponent which is half of the corresponding one for periodic systems.
On the right of Fig. \ref{1intdbc} we report the logarithmic negativity as function of the chord length $L\sin(\pi \ell/L)$
for $L=25,50,100,200$.
All the data at different $L$ collapse on the same curve that for large enough chord length are perfectly described 
by the CFT prediction (\ref{1intobcfin}).
The corrections to the scaling are smaller than the ones for  $n\geq 2$, in agreement 
with the general analysis of the R\'enyi entropies  \cite{un1,un2,fc-11}.

Finally, we checked that all the results reported in this subsection satisfy  the relations in Eq. (\ref{pureqm})
between $\Tr(\rho^{T_2})^n$ (and ${\cal E}$) and the R\'enyi entropies in the same state. 

\subsection{The negativity for two adjacent intervals in periodic chains.}

\begin{figure}[t]
\includegraphics[width=\textwidth]{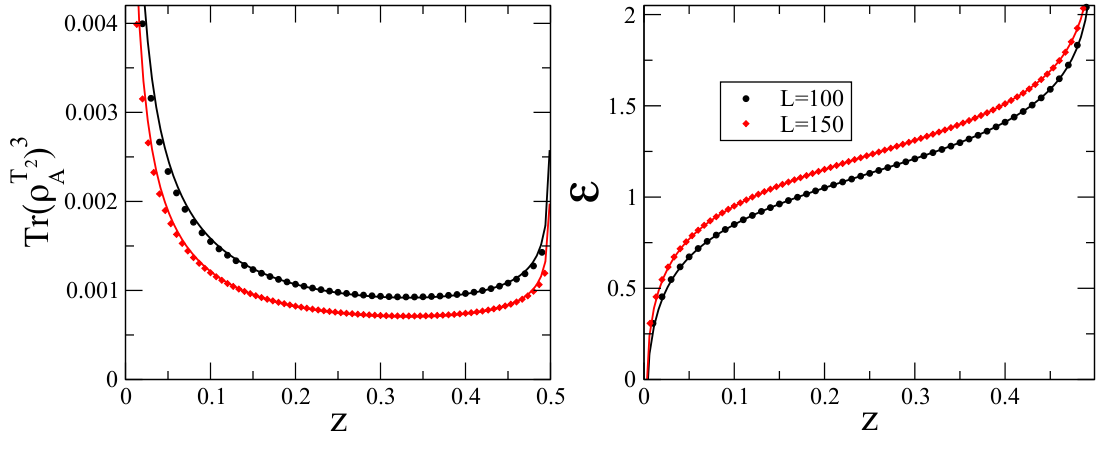}
\caption{A typical example of the entanglement of two adjacent intervals of the same length $\ell$ in a periodic chain of length $L$. 
The legend refers to both plots and we fix $\omega=10^{-6}$.
Left: $\Tr (\rho_A^{T_2})^3$ as function of $z=\ell/L$. The data are consistent with the 
conformal prediction (full lines with only one multiplicative free parameter).
Right: Logarithmic negativity ${\cal E}$ for the same chains as on the left. Again the data 
are perfectly consistent with the CFT predictions (full lines with only one additive free parameter).
}
\label{adjexem}
\end{figure}

We now consider the case of two adjacent intervals in a periodic system of total length $L$.
For simplicity we consider the two intervals to have equal length $\ell$, thus all the 
results depend on the single parameter $z\equiv \ell/L\in[0,1/2]$.
In terms of $z$, the CFT predictions in Eqs. (\ref{3ptfinL}) can be rewritten as (we fix $c=1$)
\be\fl
\Tr(\rho_A^{T_2})^n\propto \left\{
\begin{array}{ll}
( \sin(\pi z))^{-(n_e/2-2/n_e)/3} 
( \sin (2\pi z))^{-(n_e/2+1/n_e)/6}\,,
\\ \\
(\sin^2(\pi z) \sin (2\pi z))^{-(n_o-1/n_o)/12},
\end{array}
\right.
\ee
and the proportionality constants depend on $L$ in a known manner.
For the logarithmic negativity from Eq. (\ref{neg3finL}), we have 
\be
{\cal E}= \frac14 \ln \frac{\sin^2(\pi z)}{\sin(2\pi z)} +{\rm cnst}= \frac14 \ln[ \tan(\pi z)] +{\rm cnst}
\,,
\ee
where again the additive constant depends on $L$ in a known manner. 
These predictions are checked against the numerical data in Fig. \ref{adjexem}
where the only free parameters in each curve is fixed by a fit. 
The agreement is clearly excellent.

\begin{figure}[t]
\includegraphics[width=\textwidth]{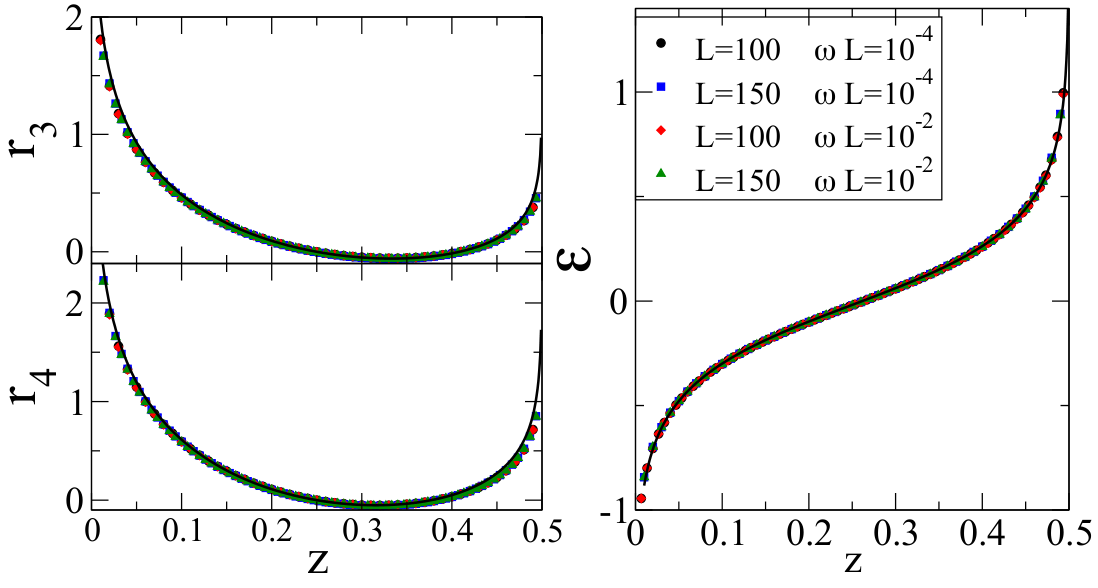}
\caption{
Entanglement for two adjacent intervals of equal length $\ell<L/2$ in a periodic chain of length $L$. 
Left: the ratio  $r_n(z)$ in Eq. (\ref{ratiorn})  as function of $z=\ell/L$
compared with the parameter free CFT prediction.
Right: Subtracted negativity $\e(z)$ in Eq. (\ref{epsdef}) 
again compared with the parameter free CFT prediction.
}
\label{3ptfig}
\end{figure}

However one can do even better checks of the theory by constructing ratios in which {\rm all}
the dependence on non-universal parameters (such as those coming from the zero mode) 
and also the dependence on $L$ cancel. 
A straightforward idea would be to divide $\Tr(\rho_A^{T_2})^n$ for the value it assumes 
at a given fixed $\ell$, e.g. $\ell=L/4$, i.e. by considering the logarithm of the ratio
\be
r_n(z)= \ln \frac{\Tr(\rho_A^{T_{A_2=\ell}})^n}{\Tr(\rho_A^{T_{A_2=L/4}})^n},
\label{ratiorn}
\ee
whose parameter free CFT predictions for $n$ even and odd are
\bea
r_{n_e}=\frac16\Big(\frac2{n_e}-\frac{n_e}2  \Big)\ln (2 \sin^2(\pi z))-
\frac16 \Big(\frac{n_e}2 +\frac1{n_e} \Big) \ln ( \sin (2\pi z))\,,
\\
r_{n_o}=
\frac1{12}\Big(\frac1{n_o}- n_o\Big)\ln (2\sin^2(\pi z) \sin (2\pi z)).
\eea
These ratios are shown in Fig. \ref{3ptfig} for $n=3$ and $n=4$. The agreement of the 
numerical data and the CFT prediction is excellent. 
Some very small deviations are visible for  $z$ close to $0$ and $1/2$, 
but we checked that they go to zero increasing $L$ in a controllable way.

For the logarithmic negativity, we can analogously define the subtracted quantity
\be
\e(z)={\cal E}(\ell,L)-{\cal E}(L/4,L)=\frac14 \ln[ \tan(\pi z)] \,,
\label{epsdef}
\ee
and again the rhs is a parameter free CFT prediction.
In Fig. \ref{3ptfig} (left panel), this prediction is compared with the numerical data 
and the agreement is extremely good since no deviations are visible even close to 
the boundaries.

\subsection{The negativity for two disjoint intervals in the periodic chain.}

\begin{figure}[t]
\includegraphics[width=.7\textwidth]{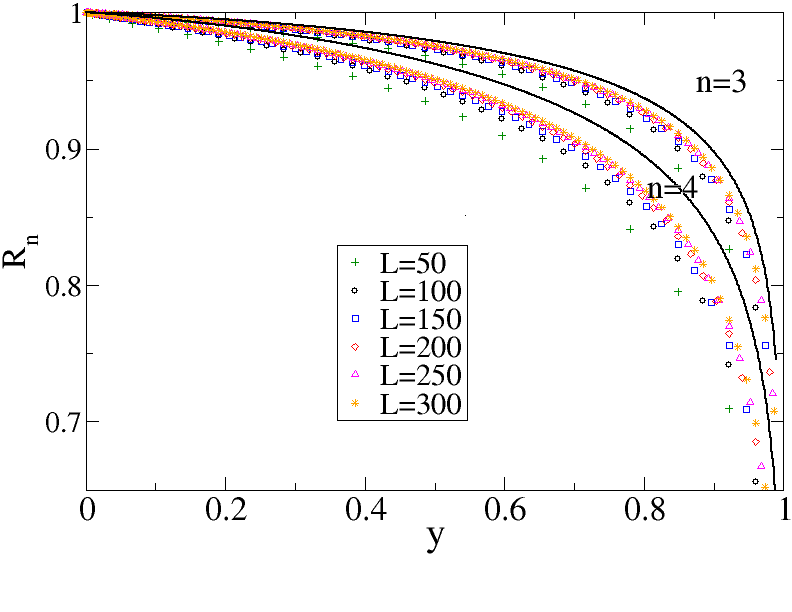}
\caption{For a periodic chain of length $L$, we report 
the ratio $R_n(y)$ defined in Eq. (\ref{Rndef}) as function of $y$ for several $L$ and for $n=3,4$.
The continuous lines are the parameter free CFT predictions to which the data converges for $L\to\infty$. 
}
\label{4ptfiga}
\end{figure}

To conclude our analysis  of the periodic chain we consider now the most difficult situation of 
two disjoint intervals for which the negativity has been already considered numerically \cite{Neg2}.
Here we start by considering the traces $\Tr(\rho_A^{T_2})^n$ which, in the conformal regime 
$\omega L\ll 1$, should be described by Eq. (\ref{Gn}) with ${\cal G}_n(y)$ given in (\ref{decomp}). 
The direct numerical data, that we do not present here, agree  well with the CFT predictions
where the overall constant is fixed by a fit and explicitly depends on the values of $\omega$ 
and $L$, since the data are influenced by the zero mode.
However, while this is a further confirmation of the predictive power of CFT, 
it gives not much information on the true entanglement (which is only obtained in the limit $n_e\to 1$).
Indeed the function ${\cal G}_n(y)$ turns out to be very close to a constant (in proper units equals 1) and
the direct data mainly probe the prefactor to ${\cal G}_n(y)$ in Eq. (\ref{Gn}),
whose logarithm vanishes in the replica limit $n_e\to1$, and so does not give any contribution to the negativity.

\begin{figure}[t]
\includegraphics[width=\textwidth]{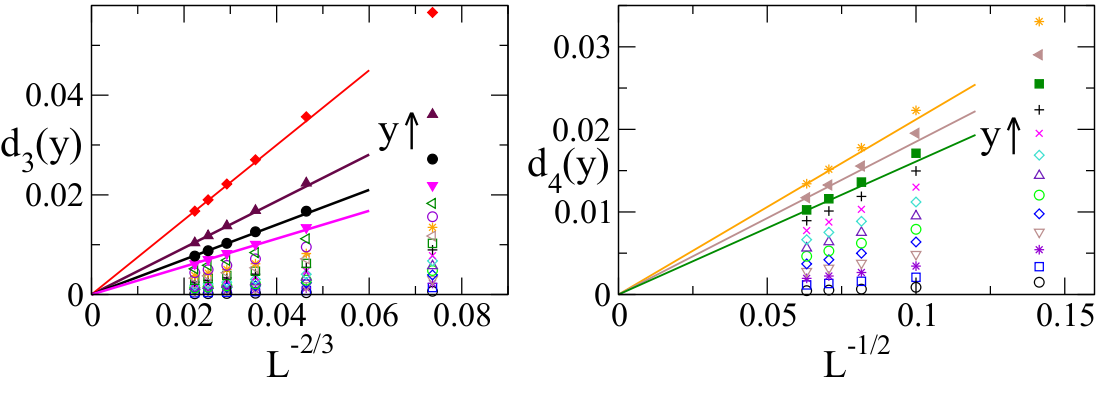}
\caption{
Finite size scaling analysis for $d_n(y)$ in Eq. (\ref{dny}) for $n=3$ (left) and $n=4$ (right).
We report  from Fig. \ref{4ptfiga} several values of $y$ (increasing in the direction of the arrow, but we do not 
give the actual value to simplify the reading of the plot). 
The data are compatible with a leading correction to the scaling of the form $L^{-2/n}$ . 
}
\label{FSS}
\end{figure}

As already stated in Sec. (\ref{neg2dis}), a practical way to get rid of the prefactor is to consider the ratio
$R_n(y)$ in Eq. (\ref{Rndef}), where also  the non-universal parts due to the zero mode cancel and we 
are left with a universal function of $y$.
The CFT prediction for this ratio $R_n^{\eta=\infty}(y)$ in Eq. (\ref{RnCFT})
 is compared to the numerical data in Fig.~\ref{4ptfiga}.
As $L$ increases, the data approach the CFT result. 
The differences with the asymptotic formula are due to the 
presence of unusual corrections to the scaling \cite{un1,un2} whose leading part is of the form $L^{-2/n}$.
A quantitative finite size scaling analysis  is reported in Fig. \ref{FSS} for $n=3,4$ 
showing that the difference  
\be
d_n(y)\equiv R_n(y)-R_n^{\eta=\infty}(y),
\label{dny}
\ee 
for several values of $y$ is of the 
expected form $L^{-2/n}$. As well known (even analytically) for other simpler cases 
\cite{un1,ce-10} for larger $n$, the subleading corrections to the scaling, 
of the form $L^{-2p/n}$ with $p$ integer, cannot be neglected and a proper 
analysis requires the introduction of some fitting parameters.

Finally we turn to the study of  the negativity ${\cal E}$ reported in Fig. ~\ref{negfig} 
showing that all data collapse on a single curve, without  sizable corrections. 
Unfortunately we do not have the analytic continuation of $R_{n_e}^{\eta=\infty}(y)$
to $n_e\to1$ as a function of $y$.
However we can study the two interesting regimes of far and close intervals 
corresponding to $y\to0$ and $y\to1^-$ respectively.
For small $y$, the data, being very close to zero, are consistent with 
the prediction that they vanish faster than any power. 
In Ref. \cite{Neg2}, on the basis of the numerical data, the form  
$e^{-a/\sqrt{y}}$ has been proposed. This proposal is shown in 
logarithmic scale on the inset of Fig. \ref{negfig} together with a simple 
exponential $e^{-b/y}$. 
The two possible scenarios are very difficult to be distinguished on the 
basis of the numerical calculations involving exponentially small numbers
and the ambiguity can be resolved only by analytically continuing $R_{n_e}^{\eta=\infty}(y)$.

For $y\to1$, the general prediction from CFT (\ref{Genclose}) for $c=1$ is 
\be
{\cal E}(y)=-\frac14 \ln (1-y) +\dots\,,
\ee
however, we have shown in Eq. (\ref{Eclose}) that for the model at hand, 
subleading double logarithmic corrections are present 
\be
{\cal E}(y)=-\frac14 \ln (1-y) - \frac12 \ln K(y) -\ln  P_1 +o(1)\,,
\label{Eexp1}
\ee
where $K(y)$ is the elliptic integral of the first kind and $P_1=0.832056\dots$ is given in Eq. (\ref{PP1}). 
Written in this form, the expansion contains also some of the subleading corrections (in $1-y$)
and it is expected to describe the data better. 
Indeed in Fig. \ref{negfig}, this prediction is almost indistinguishable from the data all the 
way from $y\sim1$ (where it is an exact result) down to $y\sim0.3$. 
We should mention that the subleading logarithmic correction may be responsible for the 
exponent $1/3$ found in Ref.~\cite{Neg2}  as compared with our analytic result $1/4$.

\begin{figure}[t]
\includegraphics[width=\textwidth]{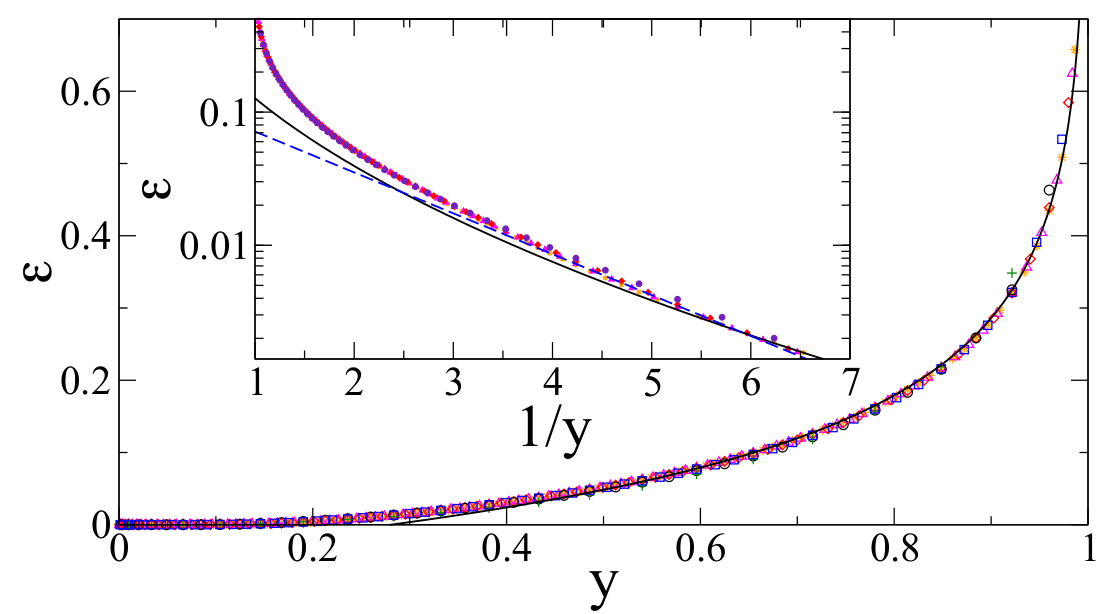}
\caption{
The negativity ${\cal E}(y)$ is a universal scale invariant function with an essential singularity at $y=0$. 
We report the data for $L=50,100,150,200,250,300$ but, since they are hardly distinguishable, 
we do not give a legend box. The solid line is the expansion close to $y\sim1$ in Eq. (\ref{Eexp1})
which very surprisingly describes well the data down to $y\sim 0.3$.
The inset shows the same plot in logarithmic scale showing that for small $y$ the two possibilities 
${\cal E}\sim e^{-a/y}$ and ${\cal E}\sim e^{-b/\sqrt{y}}$ are too close to be distinguished.
}
\label{negfig}
\end{figure}

Finally we would like to mention that, in a long enough chain, when each interval contains a {\it finite} number of lattice points, 
the negativity must vanish exactly for sufficiently large separations. 
Indeed the reduced density matrix $\rho_A$ has all strictly positive eigenvalues
and in the limit when the intervals are far apart $\rho_A$ factorizes and each factor is a finite matrix 
with positive eigenvalues which become independent of the separation.
Thus, when we take the partial transpose, the change in the density matrix, and therefore in the eigenvalues, 
can be made arbitrarily small since $\langle p_i p_j\rangle$ 
(which is the correlator that changes sign, cf. Eq. (\ref{Pt2})) decreases like $|i-j|^{-2}$.
Thus the partial transpose changes the elements of $\rho_A$ but an amount which is arbitrarily 
small so the eigenvalues do not change sign.
Indeed this is consistent with the well-known result \cite{rev,fazio} 
that the entanglement of two far away sites is exactly zero.

\subsection{Tripartite chains with Dirichlet boundary in the origin}

Now we consider the non-trivial case of a tripartite chain on a system with boundaries  
discussed in Sec. \ref{bou0}, which is the 
semi-infinite line, with $A_1=[0,\ell]$, $A_2=[\ell,2\ell]$ and $B$ the remainder. 
In this case, the results for $\Tr (\rho_A^{T_2})^n$ are given in Eq. (\ref{2adjbound})
which we report also here:
\be
\Tr (\rho_A^{T_2})^n=\left\{
\begin{array}{ll}
\ell^{-c/6(n-1/n)} & n\; {\rm odd},\\
\\
\ell^{-c/12(n-1/n) -c/6(n/2-2/n)} & n\; {\rm even}.
\end{array}
\right.
\ee
The most interesting feature is the peculiar behavior for $n$ even, which can be explicitly checked 
by considering  the semi-infinite harmonic chain with  $\omega=0$.
The results for the numerical evaluation of $\Tr (\rho_A^{T_2})^n$ are 
reported in Fig. \ref{openfig} (left), showing the perfect agreement of both even and 
odd $n=3,4,5,6$ where only the global amplitude has been fixed with a fit. 
The right panel of the same figure shows the corresponding logarithmic negativity 
which turns out to be described by the CFT prediction 
${\cal E}=1/4 \ln \ell+$ const.

\begin{figure}[t]
\includegraphics[width=\textwidth]{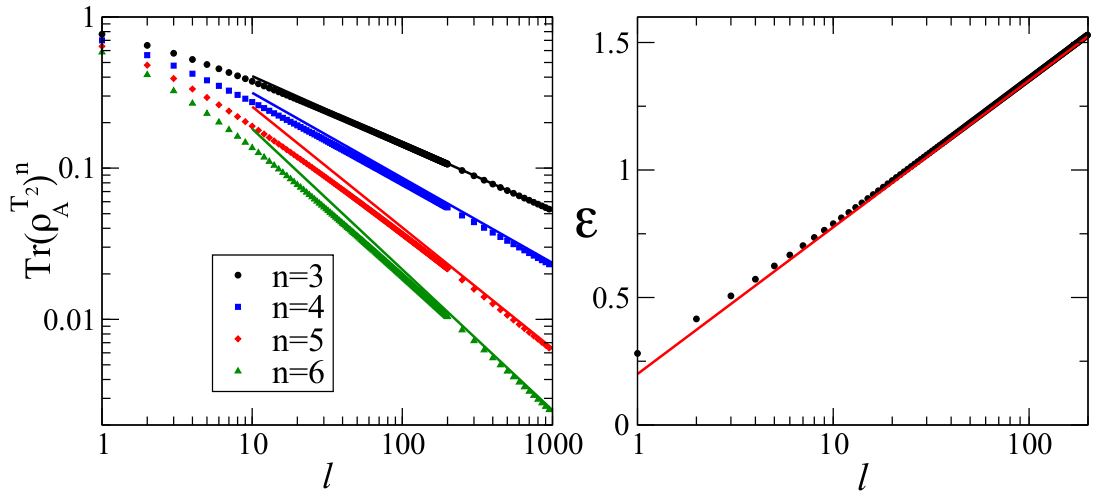}
\caption{
$\Tr (\rho_A^{T_2})^n$ (left) and logarithmic negativity (right) for a semi-infinite critical ($\omega=0$) 
harmonic chain with Dirichlet boundary conditions in the origin.
We report the results for two adjacent intervals of  equal length $l$ with the first one starting from the origin. 
The data are perfectly described by the asymptotic CFT predictions reported as solid lines. 
}
\label{openfig}
\end{figure}

\section{Some scaling considerations for massive theories}

It is well known that for a gapped one-dimensional  model,
increasing $\ell$ the entanglement (R\'enyi) entropy saturates to a finite value \cite{Vidal}.
The calculation of this saturation value is generically complicated 
because it depends on microscopical details of the model and 
so it is calculable only for few simple integrable cases (see e.g. \cite{cc-04,ijk-05,petal}).
However, some general results can be worked out 
when a system is close to a conformal quantum critical point,
and in the scaling limit where the lattice spacing
$a\to0$ and with the correlation length  fixed and large.
Under these hypotheses and  when all the lengths of the various 
subsystems (and so the total one for a finite system) are much larger than $\xi$
which is itself large, the R\'enyi entropies are \cite{cc-04}
\be
S_A^{(n)}={\cal A}\frac{c}{12}\Big(1+\frac1n\Big)\log\frac{\xi}a+O(\xi^0)\,,
\ee
where ${\cal A}$ is the number of boundary points between $A$ and its
complement and $c$ is the central charge of the conformal field theory for $\xi^{-1}=0$.  
Clearly,  when the interval lengths are of the order of
$\xi$, a complicated and universal (within the scaling limit) crossover takes place, 
which has been worked out only in very few QFT \cite{cc-04,ccd-08,somecasini,ben}.

The same difficulties for the entanglement entropy, with also some additional problems, make 
prohibitive the calculations of  the negativity and $\Tr(\rho_A^{T_2})^n$ for a 
general  gapped model. 
However, in the scaling limit and when the correlation length is itself large, but smaller 
than all the subsystems lengths,  scaling considerations lead to very general results. 
For example in the case of a bipartition of a pure state in a finite interval of length $\ell\gg\xi$
and the remainder, the scaling suggests to replace $\ell$ with $\xi$ in Eqs. (\ref{1inte}) and (\ref{1into}) 
for $\Tr(\rho_A^{T_2})^n$ and so the negativity is
\be
{\cal E}=\frac{c}2\ln \xi+O(\xi^0)\,.
\ee
This is clearly true because it is the general relation (\ref{pureqm}) between negativity and R\'enyi entropy. 
A direct check of  this is  the general result for a bisected harmonic chain
(i.e. $A_2$ formed by $N/2$ consecutive sites and $A_1$ the remainder) 
with periodic boundary conditions and nearest neighbor interaction.
In this case,  an interesting exact formula has been found for the logarithmic negativity \cite{Audenaert02}
\begin{equation}
\label{LN bisected}
{\cal E}= \frac{1}{4} \ln\bigg(1+\frac{4 K}{M \omega^2}\bigg),
\end{equation} 
which, remarkably, is independent of the size of the chain $N$.
In the limit $\xi^{-1}\propto\omega\ll 1$ gives ${\cal E}=(\ln \xi)/2$, in agreement with the general scaling.

\begin{figure}[t]
\includegraphics[width=.7\textwidth]{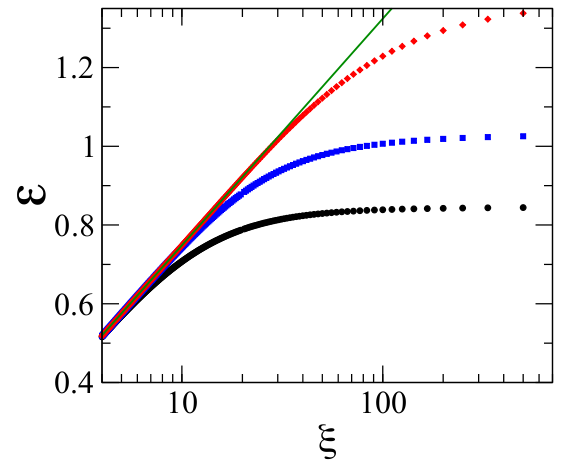}
\caption{
The negativity ${\cal E}$ for two adjacent intervals of length $\ell=10,20,50$ (from bottom to top) in 
a periodic system of total length $L=150$.
For $\xi\leq \ell$ all data collapse on the scaling prediction (\ref{e2adintmass}) while, when 
$\xi$ becomes of order $\ell$, a crossover to the CFT scaling form takes place. 
}
\label{massfig}
\end{figure}

A more interesting example  can be obtained by considering the case of two adjacent intervals.
Again, scaling suggests to substitute in Eq. (\ref{3ptevenbis}) for the conformal case $\ell_{1,2}$ with $\xi$
giving 
\be
{\cal E}=\frac{c}4\ln \xi+O(\xi^0)\,.
\label{e2adintmass}
\ee
This scaling in $\xi$ is checked in Fig. \ref{massfig} (left) for the harmonic chain for two adjacent intervals 
of the same length $\ell$. This shows an excellent agreement for $1\ll \xi\ll \ell$ while for larger $\xi$, the 
crossover to the CFT prediction (\ref{3ptevenbis}) takes place.

On the same lines as above, the negativity for other tripartitions can be deduced by means of simple
scaling arguments.

\section{Conclusions}

We introduced a general field theoretical formalism to calculate the negativity and the logarithmic negativity, as an
extension of our previous short communication \cite{us-letter}. 
This novel approach is based on a replica calculation of the traces of even integers $n_e$ powers of the partial transpose of 
the reduced density matrix and analytically continuing this to $n_e\to1$, i.e. the logarithmic negativity is
\be
{\cal E}= \lim_{n_e\to1} \ln \Tr (\rho_A^{T_2})^{n_e}\,.
\ee
Several physical situations have been explicitly worked out for a conformally invariant theory:
\begin{itemize}
\item The case in which $A_1$ is an interval and $A_2$ the remainder of an infinite, 
semi-infinite or finite (both periodic and with boundaries) system.
\item The case in which $A_1$ and $A_2$ are two adjacent intervals of length $\ell_1$ and $\ell_2$ respectively.
For an infinite system the negativity is
\be
{\cal E}=\frac{c}4 \ln \frac{\ell_1\ell_2}{\ell_1+\ell_2} +{\rm cnst}\,.
\ee
This is simply generalized to a finite periodic system. 
\item For the case in which $A_1$ and $A_2$ are two disjoint intervals (always of length $\ell_1$ and $\ell_2$), 
the negativity turns out to depend on the full operator content of the theory and it is a {\it scale invariant 
function} (i.e. a function of the harmonic ratio of the four points defining the two intervals).
We calculated the traces of integer powers of the partial transposed reduced density matrix for a free bosonic theory,  
both compactified (i.e. the Luttinger liquid field theory) and in the limit of infinite compactification radius.
However, the $n$ dependence of these formulas is too complicated and we managed to calculate the 
analytic continuation only in some physical relevant limits (far and close intervals). 
\end{itemize}
All the CFT results have been accurately checked for a chain of harmonic oscillators, finding 
perfect agreement. We also proposed a few scaling relations valid in the scaling regime 
when a theory is close to a conformal quantum critical point, and the correlation length
(inverse gap or mass) is finite but large. 

We are currently working out the generalization of the CFT approach to finite temperature 
field theories which is not as straightforward as it could naively appear. 
Finally, we  mention that it is of extreme interest to check numerically all our CFT predictions  in more 
complicated lattice models such as spin-chains and itinerant fermions described by the Luttinger liquid 
field theory, for which we worked out explicit predictions for $\Tr (\rho_A^{T_2})^{n}$
even in the case of disjoint intervals.

\section*{Acknowledgments} 
ET thanks Marcus Cramer for discussions. 
This work was supported by the 
ERC under  Starting Grant  279391 EDEQS (PC). 
This work has been partly done when all the authors were guests of the Galileo Galilei Institute in Florence 
and Institut Henri Poincar\'e in Paris.
ET thanks the Dipartimento di Fisica dell'Universit\`a di Pisa for the warm hospitality during the last part of this work.

\appendix


\end{document}